\documentclass{emulateapj}
\usepackage{amstext}
\usepackage{amsmath}
\shorttitle{Two-Component Jet Models of GRB Sources}
\shortauthors{Peng, K\"onigl, \& Granot}

\newcommand{\be}{\begin{equation}}
\newcommand{\ee}{\end{equation}}
\newcommand{\bearr}{\begin{eqnarray}}
\newcommand{\eearr}{\end{eqnarray}}
\newcommand{\rmn} {{\rm n}}
\newcommand{\rmw} {{\rm w}}
\newcommand{\rmc} {{\rm c}}
\newcommand{\rmm} {{\rm m}}
\newcommand{\rmd}{{\rm d}}
\newcommand{\rme}{{\rm e}}
\newcommand{\rmB}{{\rm B}}
\newcommand{\rmiso}{{\rm iso}}
\newcommand{\rmdec}{{\rm dec}}
\newcommand{\rmjet}{{\rm jet}}
\newcommand{\rmj}{{\rm j}}
\newcommand{\rmobs}{{\rm obs}}
\newcommand{\rmmax}{{\rm max}}
\newcommand{\bc}{\begin{center}}
\newcommand{\ec}{\end{center}}

\begin{document}

\title{Two-Component Jet Models of Gamma-Ray Burst Sources}

\author{Fang Peng,\altaffilmark{1} Arieh K\"onigl,\altaffilmark{1}
and Jonathan Granot\altaffilmark{2}}
\altaffiltext{1}{Department of Astronomy \& Astrophysics and Enrico Fermi
Institute, University of Chicago, 5640 South Ellis Avenue,
Chicago, IL 60637; fpeng@oddjob.uchicago.edu, arieh@jets.uchicago.edu.}
\altaffiltext{2}{Kavli Institute for Particle Astrophysics and Cosmology,
Stanford University, P.O. Box 20450, MS 29, Stanford, CA 94309;
granot@slac.stanford.edu.}

\begin{abstract}

  Recent observational and theoretical studies have raised the
  possibility that the collimated outflows in gamma-ray burst (GRB)
  sources have two distinct components: a narrow (opening half-angle
  $\theta_{\rmj,\rmn}$), highly relativistic (initial Lorentz factor
  $\eta_\rmn \gtrsim 10^2$) outflow, from which the $\gamma$-ray
  emission originates, and a wider ($\theta_{\rmj,\rmw} \lesssim 3\, 
\theta_{\rmj,\rmn}$),
  moderately relativistic ($\eta_\rmw\sim 10$) surrounding flow.
  Using a simple synchrotron emission model, we calculate the R-band
  afterglow lightcurves expected in this scenario and derive algebraic
  expressions for the flux ratios of the emission from the two jet
  components at the main transition times in the lightcurve.  For viewing
  angles $\theta_\rmobs < \theta_{\rmj,\rmn}$ we find that the contribution
  of the wide component to the optical afterglow is negligible if its
  kinetic energy $E_\rmw$ is significantly smaller than that of the
  narrow component, $E_\rmn$, as expected for the jet core and cocoon
  outflow components in the collapsar jet-breakout model. However, if
  $E_\rmw/E_\rmn>1$ [but the isotropic-equivalent energy ratio
  $E_{\rmiso,\rmw}/E_{\rmiso,\rmn}=(E_\rmw/E_\rmn)(\theta_{\rmj,\rmn}/
  \theta_{\rmj,\rmw})
  ^2$ remains $<1$], as expected for the decoupled neutron and
  proton components, respectively, in an initially neutron-rich,
  hydromagnetically accelerated jet model, then the narrow component
  only dominates the early afterglow and the wide component takes over
  after its nominal deceleration time $t_{\rmdec,\rmw}$ (typically
  $\sim 0.1-1\ {\rm days}$). 
  Given that $t_{\rmdec,\rmw}$ is comparable to the jet-break time
  $t_{\rmjet,\rmn}$ of the narrow component for characteristic parameter
  values, the emergence of the wide component at $t_{\rmdec,\rmw}$ may
  mask the jet break in the narrow component at $t_{\rmjet,\rmn}$,
  which in turn may lead to an overestimate of the $\gamma$-ray energy
  emitted by the source and hence of the required $\gamma$-ray emission
  efficiency. We apply this scheme also to X-ray flash sources, which
  we interpret as GRB jets viewed at an angle $\theta_\rmobs > 
  \theta_{\rmj,\rmn}$.
  Finally, we argue that a neutron-rich hydromagnetic outflow may
  naturally give rise to repeated brightening episodes in the afterglow
  lightcurve as observed in GRB 021004 and GRB 030329.

\end{abstract}

\section{Introduction}
\label{introduction}

Gamma-ray bursts (GRBs) and their afterglows are commonly interpreted
in terms of a relativistic outflow that emanates from the vicinity of
a solar-mass neutron star or black hole \citep[e.g.,][]{P99,M02}. In
this picture, the prompt gamma-ray emission is attributed to a highly
relativistic ejecta (with an initial Lorentz factor $\gamma \gtrsim
10^2$), whereas the subsequent afterglow emission in the X-ray,
optical, and radio (over hours, days, and weeks, respectively,
after the GRB) arises from the shock that is driven into the ambient
medium as the ejecta sweeps up the external medium and decelerates.
Most afterglow observations to date have been carried out
hours to days after the GRB event, by which time the Lorentz
factor of the afterglow shock has decreased to $\lesssim 10$.
These observations have revealed the presence of achromatic
breaks in the afterglow lightcurves of many sources, which
strongly indicate that GRB outflows are collimated into
narrow jets \citep{R99,SPH99}.

Recently, however, the possibility that at least some GRB outflows
consist of two distinct components has been raised in the literature.
On the observational side, this possibility was first invoked by
\citet{P98},
who suggested that the afterglow from GRB 970508
could be explained in terms of a narrow jet surrounded by an isotropic
outflow. \citet{F00} subsequently proposed that the $\gamma$-rays and
early (shorter-wavelength) afterglow emission in GRB 991216 could be
attributed to a narrow, ultrarelativistic outflow component, and that
the late (longer-wavelength) afterglow emission in this source
originates in a separate wide component that is only mildly
relativistic. A similar picture was proposed for GRB 030329 by
\citet{B03} and \citet[][]{S03}. A two-component model was also
suggested as an explanation of the observed rebrightening of the X-ray
flash (XRF) source XRF 030723 \citep[][]{H04} as well as of the apparent
peak-energy distribution of GRBs and XRFs \citep[][]{LD04}
and of the origin of the blueshifted optical absorption features in
the spectrum of the GRB 021004 afterglow \citep[][]{S05}.

The possibility of a two-component outflow in GRB sources has
been independently indicated by theoretical
considerations. One can broadly divide the physical models that
give rise to such an outflow into two classes: models in
which the separation into two components is an intrinsic property of
the outflow, and those in which a narrow relativistic jet gives rise
to a wider and slower component as it propagates through (and interacts
with) the envelope of a progenitor massive star. One example of a model
of the first type was worked out by \citet{LE93} and \citet{VPL03}: it
consists of (1) a relativistic, baryon-poor jet driven
electromagnetically along disk-anchored magnetic field lines that
thread the horizon of a rotating black hole, and (2) a
subrelativistic, baryon-rich wind that is driven thermally from that
disk. Another example is provided by hydromagnetically driven jets
that originate from a neutron star or a neutron-rich accretion disk
that form in the collapse of a massive star \citep{VPK03}. In this
case the neutrons decouple at a moderate Lorentz factor while the
protons continue to be accelerated and collimated by the
electromagnetic forces, giving rise to a narrow, highly relativistic
proton component and a wider and slower neutron component (which,
after decoupling, is transformed into a moderately relativistic proton
component through neutron decay). Examples of models of the second
type include jet-induced core-collapse supernovae, wherein collimated
high-velocity jets cause the envelope of the progenitor massive star
through which they propagate to be ejected with subrelativistic speeds
and an oblate geometry \citep{K99}, and the collapsar model, in which
the outflow resulting from the jet breakout through the progenitor
star's envelope is predicted to consist of a highly relativistic jet
core and a moderately relativistic surrounding cocoon
(\citealt{ZWH04}; see also \citealt{RCR02}).

In this paper we focus on two-component GRB outflows in which
both components initially move with relativistic speeds
and therefore end up contributing to the optical afterglow
emission. Accordingly, we adopt as representative examples the
hydromagnetically accelerated, initially neutron-rich jet model
of \citet{VPK03} and the collapsar jet-breakout model of
\citet{ZWH04}. According to the numerical simulations of the
latter authors, the narrow component in the collapsar model has
a Lorentz factor $\eta_{\rmn} \gtrsim 100$ and an opening
half-angle $\theta_{\rmj,\rmn} \sim 3^\circ-5^\circ$, whereas
the corresponding quantities for the wide component are
$\eta_{\rmw} \sim 15$ and $\theta_{\rmj,\rmw} \sim 10^\circ$, respectively.
The characteristic Lorentz factors in this scenario are very similar
to those ($\eta_{\rmn} \sim 200$ and $\eta_{\rmw} \sim 15$) in the
representative neutron-rich hydromagnetic model of
\citet{VPK03}.\footnote{In the simplified model used by \citet{VPK03},
the value of $\theta_{\rmj,\rmw}$ could not be calculated exactly; in
this paper we assume that it can be as large as $\sim
3~\theta_{\rmj,\rmn}$.} However, {\em in contrast} with the collapsar
model, in which the highly relativistic jet component is in general
more energetic (typically by about an order of magnitude) than the
cocoon material, the asymptotic kinetic energy $E_{\rmw}$ of the wide
component in the hydromagnetic jet model typically exceeds the
corresponding energy of the narrow component ($E_{\rmw} \approx
2\, E_{\rmn}$ in the fiducial model of \citealt{VPK03}).

Our goal in the present work is to examine some of the
observational properties of two-component GRB outflows. In
particular, we calculate (\S~\ref{AG_LC}) the approximate optical
afterglow lightcurves that are produced by the shocks that
the two jet components would drive into the ambient medium. We then argue
(\S~\ref{apply}) that outflows of this type may have significant general
implications to our understanding of GRB and XRF
sources. Our conclusions are given in \S~\ref{conclude}.

\section{Model Afterglow Lightcurves}
\label{AG_LC}

A simple jet structure is assumed in this work, consisting of a narrow
and initially faster component and a wide and initially slower
component. Each component is assumed to be uniform within some finite
opening angle and to have sharp edges.
This two-component model should not be regarded as a limiting case
of the structured ``universal'' jet models discussed in the
literature \citep[e.g.,][]{RLR02,ZM02,Z04}. In the latter models,
all jets are nearly identical and their injected energy per unit
solid angle has a power-law or a Gaussian dependence on the polar
angle $\theta$ (measured with respect to the jet axis). In contrast,
in the scenario that we consider the wide component does not
contribute to the $\gamma$-ray emission and hence cannot be a part
of a traditional universal-jet model.
In the uniform, sharp-edged jet picture the opening angles of
the two outflow components are not invariant from source to
source although their values may well be correlated. It is, however,
conceivable that
each of the two components is structured and that the jet is
``universal'' in the sense that this structure
varies little from source to source.
The basic implications of a structured two-component jet model
would be similar to the ones that we discuss in
\S~{\ref{apply} in the context of a sharp-edged, uniform
outflow, although some of the details may be different and would
depend on the specifics of the angular distribution of $E$ and
$\eta$ in each component (see, e.g., \citealt{KG03} and \citealt{Z04}).

In our simple treatment the interaction between the two jet
components is neglected. This assumption can be justified even 
after the narrow component's jet-break time (see
eq. [\ref{t_jet1}] below), when the effects of sideways
expansion could in principle become relevant
\citep[e.g.,][]{R99}, in view of indications from recent numerical
simulations \citep[e.g.,][]{KG03,CGV04} that in practice there is
relatively little lateral spreading so long as the jet is at
least moderately relativistic.
To further simplify the discussion we only consider the case of
a uniform external medium (of number density 
$n=n_0\, {\rm cm}^{-3}$) and we neglect the possible effects of
radiative losses on the hydrodynamic evolution (which may affect the
early afterglow during the period of fast cooling if the
fraction $\epsilon_e$ of the
internal energy in electrons behind the afterglow shock
is not $\ll 1$). The narrow and fast jet component has an initial
Lorentz factor $\eta_\rmn$ ($\gtrsim 10^2$), a half-opening angle
$\theta_{\rmj,\rmn}$, and a kinetic energy (at the beginning of the
afterglow phase) $E_\rmn$, while the wide and slow jet component is
characterized by $\eta_\rmw$ ($\sim 10$), $\theta_{\rmj,\rmw}$
($>\theta_{\rmj,\rmn}$), and $E_\rmw$. In the following, the subscripts
`n' and `w' will denote the narrow and wide jet components,
respectively. The ratio of the true energy $E$ and the
isotropic-equivalent energy $E_{\rm iso}$ is given by the beaming
factor $f_b=1-\cos\theta_\rmj\approx \theta_\rmj^2/2$.  Thus, 
$E_{\rm iso,w}/E_{\rm iso,n} = (\theta_{\rmj,\rmn}/\theta_{\rmj,\rmw})^2
E_\rmw/E_\rmn$.

The emission from each outflow component is calculated separately. For
the early afterglow (while the reverse shock is still present)
we use the results of \citet{SP99a,SP99b}. For the emission during the
subsequent self-similar evolution \citep{BM76} we follow
\citet{SPN98}, and for the post--jet-break emission we use the results
of \citet{SPH99}.  The typical synchrotron frequency $\nu_m$, the
cooling frequency $\nu_c$, and the peak flux $F_{\nu,{\rm max}}$, of
the emission from the shocked external medium behind the forward shock
are given by \bearr{} \nu_\rmm & = & 1.1 \times
10^{19}\,g^2\epsilon_{\rme,-1}^2
\epsilon_{\rmB,-1}^{1/2} n_0^{1/2} (\gamma/300)^4 \;{\rm Hz}\ , \label{nu_m}\\
\nu_\rmc & = & 1.1 \times 10^{17}\,\epsilon_{\rmB,-1}^{-3/2}
n_0^{-3/2}t_{\rm s}^{-2} (\gamma/300)^{-4}\;{\rm Hz}\ , \label{nu_c}\\
\label{F_numax1}
F_{\nu,\rmmax} & = & 220\,\epsilon_{\rmB,-1}^{1/2}
n_0^{3/2} D_{{\rm L},28}^{-2} t_{\rm s}^3 (\gamma/300)^8\;{\rm \mu Jy}
\eearr \citep{SP99b}, where $g\equiv 3(p-2)/(p-1)$, $p$ is the
power-law index of electron energy distribution
($dN_e/d\gamma_e\propto\gamma_e^{-p}$), $t=t_{\rm s}\;$sec is the
observed time, $\gamma$ is the Lorentz factor of the shocked
fluid, $\epsilon_B$ is the fraction of the internal energy behind the shock
in the magnetic field, $D_{\rm L}$ is the the luminosity
distance, and $Q_i\equiv Q/$($10^i$ times the c.g.s. units of $Q$).

The interaction of the jet with the ambient medium initially drives
a reverse shock into the GRB ejecta, which decelerates the ejecta. When
the reverse shock is Newtonian, or at most mildly relativistic, then
$\gamma\approx\eta$ over its entire duration and the energy given to
the swept-up external medium (of rest mass $M$) is
$\gamma^2Mc^2\sim\eta^2Mc^2$. As this energy approaches $E$, the
original kinetic energy of the ejecta, after \be\label{t_dec} t_{\rm
dec}=\frac{R_{\rm dec}}{2c\eta^2}= 0.49\,\left(\frac{E_{\rm
iso,52}}{n_0}\right)^{1/3}
\left(\frac{\eta}{10}\right)^{-8/3}\;{\rm days}\ , \ee significant
deceleration must occur. For $t>t_{\rm dec}$ most of the energy is in
the shocked external medium and a self-similar evolution is
established \citep{BM76}. Since $\eta_{\rm w}$ is assumed to be rather
small ($\sim 10$), $t_{\rm dec,\rmw}$ ($\sim 0.5\;$days) is much
larger than the duration of the GRB. Therefore, the ejecta is always
in the ``thin shell'' regime \citep{SP95,Sari97}. In this case the
reverse shock is initially Newtonian. It is natural to expect some
variation in the initial Lorentz factor, $\Delta\eta\sim\eta$. For a
thin shell, this causes the ejecta shell to start spreading long
before the reverse shock finishes crossing the shell, which in turn
causes the reverse shock to become mildly relativistic before the
crossing has ended. In this paper we concentrate on the optical
emission from the forward shock, which in the case of the fast
component would dominate the optical flux from the reverse shock after
$\sim 10^3\ {\rm sec}$ for typical parameters \citep{KZ03}.
In the case of the slow component, the contribution from the forward
shock should typically dominate the optical flux at all times, with
the emission from the reverse shock making a significant contribution
only in the radio band \citep[e.g.,][]{PNG04}.

When $\gamma$ drops to $\sim\theta_\rmj^{-1}$ the edge of the jet
becomes visible and sideways expansion may become noticeable. These two
effects cause a break in the lightcurve at
\be
\label{t_jet1}
t_{\rm jet} = 0.25 
\,E_{\rmiso,52}^{1/3}n_0^{-1/3}\theta_{j,-1}^{8/3}\;{\rm days}
\ ,
\ee
with the former effect evidently responsible for most of the
steepening if the Lorentz factor is not too close to 1.
Expressed in terms of the true energy, the jet-break time is
\be
\label{t_jet2}
t_{\rm jet} = 0.66\,E_{51}^{1/3}n_0^{-1/3}
\theta_{j,-1}^{2}\;{\rm days}\ .
\ee

In the early afterglow, at $t<t_{\rm dec}$, $\gamma\approx\eta$
\citep{SP99a,SP99b}. At $t_\rmdec < t \leq t_\rmjet$,
$\gamma\sim\theta_\rmj^{-1}(t/t_{\rm jet})^{-3/8}$ \citep{SPN98}.
At $t>t_\rmjet$ (and before the nonrelativistic transition time
$t_{\rm NR}$), we have $\gamma\sim\theta_\rmj^{-1}(t/t_\rmjet)^{-1/2}$
assuming rapid lateral expansion 
\citep{SPH99}.
Therefore, the temporal scalings of the break frequencies and peak
flux are given by
\bearr{}
\label{nu_m2}
\nu_\rmm &
\propto & \gamma^4 \propto \left\{\begin{array}{ll}
                    t^0, & t < t_\rmdec\ ,\\
                    t^{-3/2}, & t_\rmdec < t < t_\rmjet\ , \\
                    t^{-2},  &  t > t_\rmjet\ ,
                                           \end{array}
                                     \right. \\
\label{nu_c2}
\nu_\rmc &
\propto & \gamma^{-4}t^{-2} \propto \left\{\begin{array}{ll}
                    t^{-2}, & t < t_\rmdec\ ,\\
                    t^{-1/2}, & t_\rmdec < t < t_\rmjet\ , \\
                    t^0  &  t > t_\rmjet \ ,
                                                    \end{array}
                                              \right. \\
\label{F_numax2}
F_{\nu,{\rm max}} &
\propto & \gamma^8 t^3 \propto \left\{\begin{array}{ll}
                    t^3, & t < t_\rmdec\ ,\\
                    t^0, & t_\rmdec < t < t_\rmjet\ , \\
                    t^{-1},  &  t > t_\rmjet \ .
                                                    \end{array}
                                              \right.
\eearr
In the limit of negligible sideways expansion after the jet-break time,
the time dependence of $\gamma$ does not change at $t_\rmjet$, so the
behavior described by the second line in equations (\ref{nu_m2}) and
(\ref{nu_c2}) continues to hold also for $t>t_\rmjet$.  However, the
maximum flux in this case is still reduced by a factor $(\theta_\rmj
\gamma)^2$ (representing the ratio of the jet area to the beaming cone
area) as $t$ increases above $t_\rmjet$, so $t^{-1}$ in the last line of
equation (\ref{F_numax2}) is replaced by $t^{-3/4}$. For practical
applications the qualitative behavior of the lightcurve in this limit is
very similar to that in the limit of rapid lateral expansion, and since
the expressions given in equations (\ref{nu_m2})--(\ref{F_numax2}) are
the ones commonly used in the literature, we continue to employ them
in this work.

The transition time $t_0$ from fast cooling to slow cooling (when
$\nu_m=\nu_c$), and the times $t_m$ and $t_c$ when $\nu_m$ and
$\nu_c$, respectively, pass by the observed frequency $\nu$, are given
by \citep{SPN98}
\bearr{}
t_0 & = & 0.55\;g^2 E_{\rmiso,52} \epsilon_{\rme,-1}^2
\epsilon_{\rmB,-1}^2 n_0 \;{\rm hr}\ , \\
t_\rmm & = & 0.36 \; g^{4/3}E_{\rmiso,52}^{1/3}
\nu_{15}^{-2/3} \epsilon_{\rme,-1}^{4/3}
\epsilon_{\rmB,-1}^{1/3}\;{\rm hr}\ , \\
t_\rmc & = & 0.17\; E_{\rmiso,52}^{-1}
\nu_{15}^{-2}\epsilon_{\rmB,-1}^{-3} n_0^{-2}  \;{\rm hr}\ .\label{t_c}
\eearr

The transition frequency, $\nu_0$,
defined by $\nu_\rmm(t_0) = \nu_\rmc(t_0)$, is given by
\be
\nu_0 = 5.5 \times 10^{14}  g^{-1}E_{\rmiso,52}^{-1}
\epsilon_{\rme,-1}^{-1} \epsilon_{\rmB,-1}^{-5/2} n_0^{-3/2}\;{\rm Hz}\ .
\ee

These transition times together with $t_\rmdec$ and $t_\rmjet$
separate the time domain into several segments. At each time
segment, the flux is derived by comparing the observed frequency to
$\nu_\rmm$ and $\nu_\rmc$ to determine the appropriate spectral behavior;
one also makes sure to use the correct dynamical behavior for the given
time segment. The explicit expressions for the flux
contributions of the two outflow components at the different
time segments are presented in Appendix~\ref{afterglow}.

The ratio of the deceleration times of the two jet components is
\be
\frac{t_{\rmdec,\rmw}}{t_{\rmdec,\rmn}} =
\left(\frac{E_{\rm iso,w}}{E_{\rm iso,n}}\right)^{1/3}
\left(\frac{\eta_\rmw}{\eta_\rmn}\right)^{-8/3}\ , \ee which for
$\eta_n/\eta_w \sim 10$ is $\sim 10^3$. The deceleration time of
the slow component is much larger than that of the fast component,
so a bump would show up in the decaying lightcurve of the fast
component due to the emission of the slow component if
$F_{\nu,\rmw} > F_{\nu,\rmn}$ at $t=t_{\rmdec,\rmw}$.

The flux ratio $(F_{\nu,\rmw}/F_{\nu,\rmn})_{t=t_{\rmdec,\rmw}}$
depends on whether the slow/wide component decelerates
before the jet-break time of the fast/narrow component or not
(i.e., on the relative ordering of $t_{\rm dec,\rmw}$ and $t_{\rm
jet,n}$), since $\gamma$ has different time evolution indices before and
after jet break. From equations (\ref{t_dec}) and
(\ref{t_jet1}), the ratio of these two times is given by
\be\label{t_ratio}
\frac{t_{\rmdec,\rmw}}{t_{\rmjet,\rmn}} \approx
A_1\left(\frac{E_{\rm iso,w}}{E_{\rm
iso,n}}\right)^{1/3}(\eta_\rmw\theta_{\rmj,\rmn})^{-8/3}\ ,
\ee
where $A_1= A_2 C_{\rm jet}^{8/3}$ and
$A_2\equiv C_t(t>t_{\rm dec})/C_t(t_{\rmdec})\sim 1$
(with the coefficients $C_{\rmjet}$ and $C_t$
defined from $\gamma(t_{\rm jet})=C_{\rm jet}/\theta_\rmj$ and
$t=R/C_t\gamma^2c$). In this work we assume
$C_{\rm jet}=1$, $C_t(t_{\rm dec})=2$, and $C_t(t>t_{\rm dec})=4$,
for which $A_1=A_2=2$.
However, in the figures of this paper we calculate the flux by using
$C_t = 4$ at all times to ensure continuity of the plotted lightcurves;
our algebraic expressions for the component flux ratios may
therefore yield values that differ somewhat (by a factor
$\lesssim 2$) from those implied by the presented figures.

From equations (\ref{F_numax1}), (\ref{t_dec}), and the scaling of $\gamma$
with time, the peak-flux ratio of the two components at $t_{\rmdec,\rmw}$ is
\bearr{}
\left.
\frac{F_{\nu,\rmmax,\rmw}}{F_{\nu,\rmmax,\rmn}} \right|_{t=t_{\rmdec,\rmw}}
& = & \left\{\begin{array}{ll}
         \frac{E_{\rmiso,\rmw}}{E_{\rmiso,\rmn}}
           &  t_{\rmdec,\rmw} < t_{\rmjet,\rmn}\ , \\
         \left(\frac{E_{\rmiso,\rmw}}{E_{\rmiso,\rmn}}\right)
          \left(\frac{t_{\rmdec,\rmw}}{t_{\rmjet,\rmn}}\right)
           &  t_{\rmdec,\rmw} > t_{\rmjet,\rmn}\ .    \\
           \end{array}
  \right.
\label{fmax_dec}
\eearr

The flux ratio under consideration also depends on the frequency
range within which the observed frequency $\nu$ is
located at $t_{\rmdec,\rmw}$ for each of the two outflow components.
For typical parameter values, $t_{\rmdec,\rmw} > t_{0,\rmn}$, so
the narrow/fast component is in the slow cooling regime,
$\nu_{\rmc,\rmn}(t_{\rmdec,\rmw})>\nu_{\rmm,\rmn}(t_{\rmdec,\rmw})$.
We also expect the wide/slow component to be slowly cooling (see
eq. [\ref{nuc_num}]).

Based on the discussion in Appendix~\ref{afterglow}, the R-band
observation frequency $\nu_{\rm R}=5 \times 10^{14}\;{\rm Hz}$
exceeds both
$\nu_{\rmm,\rmw}(t_{\rmdec,\rmw})$ and
$\nu_{\rmm,\rmw}(t_{\rmdec,\rmn})$ for typical parameter values.
However, $\nu_{\rm R}$ can be either larger or smaller than $\nu_\rmc$
for both components. There are thus four relevant cases: \newline
If $ \nu_\rmm(t_{\rmdec,\rmw}) < \nu <
\nu_\rmc(t_{\rmdec,\rmw})$ for both components, then
\bearr{}
\left. \frac{F_{\nu,\rmw}}{F_{\nu,\rmn}}
\right|_{t=t_{\rmdec,\rmw}} & = &
\frac{F_{\nu,\rmmax,\rmw}}{F_{\nu,\rmmax,\rmn}}
\left(\frac{\nu_{\rmm,\rmw}}{\nu_{\rmm,\rmn}}\right)^{(p-1)/2}
\equiv f_1
\nonumber\\
& = & \left\{\begin{array}{ll}
         A_2^{3(p-1)/4}
         \left(\frac{E_{\rmiso,\rmw}}{E_{\rmiso,\rmn}}\right)^{(p+3)/4} & \\
           & \hspace{-1.5cm} t_{\rmdec,\rmw} < t_{\rmjet,\rmn}\ , \\ \\
          A_1^{(p+3)/4} A_2^{3(p-1)/4}
         \left(\frac{E_{\rm iso,w}}{E_{\rm iso,n}}\right)^{(p+3)/3} & \\
         \hspace{2mm} \times\,(\eta_w\theta_{\rm j,n})^{-2(p+3)/3}
           & \hspace{-1.5cm} t_{\rmdec,\rmw} > t_{\rmjet,\rmn}\ .
           \end{array}
  \right.
\label{f1}
\eearr
This parameter regime applies to all the numerical examples presented in
\S~\ref{apply}.
If $ \nu > \nu_\rmc(t_{\rmdec,\rmw})  >
\nu_\rmm(t_{\rmdec,\rmw})$ for both components, then
\bearr{}
\left.
\frac{F_{\nu,\rmw}}{F_{\nu,\rmn}} \right|_{t=t_{\rmdec,\rmw}} & =
& \frac{F_{\nu,\rmmax,\rmw}}{F_{\nu,\rmmax,\rmn}}
\left(\frac{\nu_{\rmc,\rmw}}{\nu_{\rmc,\rmn}}\right)^{1/2}
\left(\frac{\nu_{\rmm,\rmw}}{\nu_{\rmm,\rmn}}\right)^{(p-1)/2}
\equiv f_2
\nonumber \\
& = & \left\{\begin{array}{ll}
         A_2^{3(p-2)/4}
         \left(\frac{E_{\rmiso,\rmw}}{E_{\rmiso,\rmn}}\right)^{(p+2)/4} & \\
           & \hspace{-1.5cm} t_{\rmdec,\rmw} < t_{\rmjet,\rmn}\ , \\ \\
         A_1^{(p+2)/4} A_2^{3(p-2)/4}
         \left(\frac{E_{\rmiso,\rmw}}{E_{\rmiso,\rmn}}\right)^{(p+2)/3} & \\
         \hspace{2mm}\times\, (\eta_w\theta_{\rm j,n})^{-2(p+2)/3}
           & \hspace{-1.5cm} t_{\rmdec,\rmw} > t_{\rmjet,\rmn}\ .    \\
           \end{array}
  \right.
\label{f2}
\eearr
If $\nu > \nu_{\rmc,\rmn}(t_{\rmdec,\rmw})$,
$\nu_{\rmm,\rmw}(t_{\rmdec,\rmw}) < \nu <
\nu_{\rmc,\rmw}(t_{\rmdec,\rmw})$, then
\be \left. \frac{F_{\nu,\rmw}}{F_{\nu,\rmn}}
\right|_{t=t_{\rmdec,\rmw}}
= f_1 \left(\frac{\nu}{\nu_{\rmc,\rmn}}\right)^{1/2} =
f_2 \left(\frac{\nu}{\nu_{\rm c,w}}\right)^{1/2}  \ .
\ee
If $\nu_{\rmm,\rmn}(t_{\rmdec,\rmw}) < \nu <
\nu_{\rmc,\rmn}(t_{\rmdec,\rmw})$, $\nu >
\nu_{\rmc,\rmw}(t_{\rmdec,\rmw})$, then \be \left.
\frac{F_{\nu,\rmw}}{F_{\nu,\rmn}} \right|_{t=t_{\rmdec,\rmw}}
= f_1 \left(\frac{\nu}{\nu_{\rmc,\rmw}}\right)^{-1/2} =
f_2 \left(\frac{\nu}{\nu_{\rm c,n}}\right)^{-1/2}  \ .
\label{f_end}
\ee

We find that, at $t_{\rmjet,\rmw}$, $\nu_{\rm R}> \nu_\rmm$
and $\nu_\rmc > \nu_\rmm$ for both components over most of the
characteristic parameter ranges (see Appendix~\ref{afterglow}).
Given also that the flux evolution after the jet break is the same
for both components, it follows that the flux ratio after the
two components have undergone a jet break is the same as that at
$t_{\rmjet,\rmw}$.
From equations (\ref{F_numax1}), (\ref{t_jet1}), and (\ref{F_numax2}),
the peak-flux ratio of the two components at $t \geq t_{\rmjet,\rmw}$
is \footnote{In the simple jet model that we use \citep{R99,SPH99},
  the jet dynamics becomes independent of its initial opening angle at
  $t>t_{\rm jet}$ \citep{G02}. Therefore, the lightcurves at any
  given viewing angle $\theta_{\rm obs}$ depend only on the true
  energy $E$ of the outflow (in addition to the ambient density and
  microphysical parameters $\epsilon_e$ and $\epsilon_B$, which are
  assumed to be the same for the two jet components). Thus, the flux
  ratios at a given observed time and frequency after both jet
  breaks (and both deceleration times) depend only on the ratio of
  their true energies, $E_\rmw/E_\rmn$.}
\be
\left.
\frac{F_{\nu,\rmmax,\rmw}}{F_{\nu,\rmmax,\rmn}}
\right|_{t \geq t_{\rmjet,\rmw}}
=  \left(\frac{E_{\rmiso,\rmw}}{E_{\rmiso,\rmn}} \right)
\left(\frac{t_{\rmjet,\rmw}}{t_{\rmjet,\rmn}} \right)
= \left(\frac{E_\rmw}{E_\rmn} \right)^{4/3}\ .
\label{fmax_tjet}
\ee
Therefore (specializing to
the case $t_{\rmdec,\rmw} > t_{\rmjet,\rmn}$), the flux ratio at
$t_{\rmjet,\rmw}$ assumes the same form as the flux ratio at
$t_{\rmdec,\rmw}$ after $\eta_\rmw$ is replaced by
$1/\theta_{\rmjet,\rmw}$ and all the characteristic frequencies are
evaluated at $t_{\rmjet,\rmw}$. Denoting by $\tilde{f}_i$ the same
flux ratios as $f_i$ for $i=1,\ 2$, but evaluated at $t_{\rmjet,\rmw}$
instead of at $t_{\rm dec,w}$, we obtain the following simple results:
\bearr{}
\tilde{f_1} \equiv \left.\frac{F_{\nu,\rmw}}
{F_{\nu,\rmn}}\right|^{\nu<\nu_\rmc}_{t=t_{\rmjet,\rmw}}
          & = & \left (\frac{E_\rmw}{E_\rmn}\right )^{(p+3)/3}
\label{f1_tilde}
\eearr
and
\bearr{}
\tilde{f_2} \equiv \left.\frac{F_{\nu,\rmw}}
{F_{\nu,\rmn}}\right|^{\nu>\nu_\rmc}_{t=t_{\rmjet,\rmw}}
          & = & \left (\frac{E_\rmw}{E_\rmn}\right )^{(p+2)/3}\ .
\label{f2_tilde}
\eearr

Another interesting quantity is the ratio of the fluxes from the
two outflow components at the corresponding jet-break times,
$\hat{f}\equiv F_{\nu,\rmw}(t_{\rmjet,\rmw})/
F_{\nu,\rmn}(t_{\rmjet,\rmn})$.
We find
\bearr{}
\hat{f_1} \equiv \left.\frac{F_{\nu,\rmw}(t_{\rmjet,\rmw})}
{F_{\nu,\rmn}(t_{\rmjet,\rmn})}\right|_{\nu<\nu_\rmc}
          & = & \left (\frac{E_\rmw}{E_\rmn}\right )
\left (\frac{\theta_{\rmj,\rmw}}{\theta_{\rmj,\rmn}}\right )^{-2p}
\label{f1_hat}
\eearr
and
\bearr{}
\hat{f_2} \equiv \left.\frac{F_{\nu,\rmw}(t_{\rmjet,\rmw})}
{F_{\nu,\rmn}(t_{\rmjet,\rmn})}\right|_{\nu>\nu_\rmc}
          & = & \left (\frac{E_\rmw}{E_\rmn}\right )^{2/3}
\left (\frac{\theta_{\rmj,\rmw}}{\theta_{\rmj,\rmn}}\right )^{-2p}\ .
\label{f2_hat}
\eearr

So far the calculations correspond to on-axis observers, i.e.,
$\theta_\rmobs = 0$. The results, however, apply to the entire
range of observation angles
$\theta_\rmobs \lesssim 1.5\, \theta_\rmj$, as demonstrated with
the help of more realistic jet models \citep{G02}. For off-axis
observers with viewing angle $\theta_\rmobs \gtrsim
1.5\ \theta_\rmj$ the afterglow emission peaks at $t_\theta$,
the time when $\gamma \simeq \theta_\rmobs^{-1}$:
\be
t_\theta = B \left (\frac{\theta_\rmobs}{\theta_\rmj} \right )^2 t_{\rmjet}
\label{t_theta}
\ee
\citep{NPG02}, where the model-dependent factor $B$ is of the
order of unity (we adopt $B=1$ for the numerical
estimates in this work).
The lightcurve for $t>t_\theta$ is similar to the on-axis lightcurve.
The maximum flux at $t_\theta$ strongly decreases with increasing
viewing angle, as $F_{\rmmax,\nu}(\theta_\rmobs) \propto
\theta_{\rmobs}^{-2p}$ (see eqs. [9] and [10] of \citealt{NPG02}).
As we discuss below, for
$ 1.5\, \theta_{\rmj,\rmn} \lesssim \theta_\rmobs \lesssim
1.5\, \theta_{\rmj,\rmw}$ it is possible for the wide component to
be visible first and for the narrow component to become dominant
(at least temporarily) later.
The estimate of equation (\ref{t_ratio}) is now replaced by
\be
\frac{t_{\rmdec,\rmw}}{t_{\theta,\rmn}} =
\frac{A_1}{B}
\left (\frac{E_\rmw}{E_\rmn}\right )^{1/3}
(\frac{\theta_{\rmj,\rmw}}{\theta_{\rmobs}})^{2}
(\theta_{\rmj,\rmw} \eta_\rmw)^{-8/3}\ .
\label{t_ratio1}
\ee The emission from the wide and the narrow outflow components peaks
at $t_{\rmdec,\rmw}$ and $t_{\theta,\rmn}$, respectively. In
evaluating the effect of these peaks on the overall lightcurve, it is
useful to consider two flux ratios: the quotient $f^a$ of the first
and second peak components, given by
$F_{\nu,w}(t_{\rmdec,w})/F_{\nu,n}(t_{\theta,\rmn})$ if
$t_{\theta,\rmn} > t_{\rmdec,\rmw}$ and by its inverse if
$t_{\theta,\rmn} < t_{\rmdec,\rmw}$, and the ratio $f^b$
of the primary and secondary flux contributions at the time
$t = \max\{t_{\rmdec,\rmw},t_{\theta,\rmn}\}$ of the second
peak. The ratio $f^a$ determines
which component dominates the overall lightcurve, whereas $f^b$
indicates whether the secondary component can play a role in the late
afterglow. As is the case at $t_{\rmjet,\rmw}$, typically the R-band
observation frequency at $t_{\theta,\rmn}$ exceeds $\nu_\rmm$ and
$\nu_\rmc>\nu_\rmm$ for both outflow components.

If $t_{\rmdec,\rmw} > t_{\theta,\rmn}$ then $f^b$ (which in this case is
the flux ratio of the wide component to the narrow component
at $ t_{\rmdec,\rmw}$) is the same as the ratio $f$ obtained in the
on-axis case for $t_{\rmdec,\rmw} > t_{\rmjet,\rmn}$.
The difference from the on-axis case is that the flux of the narrow
component peaks at a later time ($t_{\theta,\rmn}$) and is smaller.
In this case
\be
f^a = f^{b}
\left (\frac{t_{\theta,\rmn}}{t_{\rmdec,\rmw}}\right)^{p}
\label{fa}
\ee
or, written down explicitly,
\bearr{}
f_1^a & = & B^p \left (\frac{A_2}{A_1}\right )^{3(p-1)/4}
\left (\frac{E_\rmw}{E_\rmn}\right)
\left (\frac{\theta_{\rmj,\rmw}}{\theta_\rmobs}\right)^{-2p} \nonumber\\
& & \hspace{2.5cm} \times\,(\theta_{\rmj,\rmw} \eta_\rmw)^{2(p-1)} \ ,
\label{fa_1}
\eearr
\bearr{}
f_2^a & = & B^p \left (\frac{A_2}{A_1}\right )^{3(p-2)/4}
\left (\frac{E_\rmw}{E_\rmn}\right )^{2/3}
\left (\frac{\theta_{\rmj,\rmw}}{\theta_\rmobs}\right)^{-2p} \nonumber\\
 &  & \hspace{2.5cm} \times\,(\theta_{\rmj,\rmw} \eta_\rmw)^{2(3p-2)/3}  \ ,
\label{fa_2}
\eearr
for the cases where $\nu < \nu_\rmc$ and $\nu > \nu_\rmc$, 
respectively.
In order for a bump associated with the wide component to become visible
during the late afterglow it is necessary for $f^b$ to exceed 1.
This condition requires
$E_\rmw/E_\rmn > A_1^{-3/4}A_2^{-9(p-2)/4(p+2)}(\eta_w\theta_{\rmj,\rmw})^2$
(see eqs. [\ref{f1}] and [\ref{f2}]).
If $f^{a}$ is also $>1$ then the wide component dominates the entire
lightcurve, with the narrow component possibly becoming visible as a
bump on the curve's rising branch. If, however, both $f^{a}$ and
$f^b$ are $<1$ then the narrow component would dominate at all times.

When $t_{\rmdec,\rmw}< t_{\theta,\rmn} < t_{\rmjet,\rmw}$, 
$f^b$ is given by
\bearr{}
f_1^b & \equiv & \left.\frac{F_{\nu,\rmn}}{F_{\nu,\rmw}}
\right|_{t=t_{\theta,\rmn}}^{\nu < \nu_c} \nonumber \\
      & = & B^{-(p+3)/4} \left (\frac{E_\rmn}{E_\rmw}\right )^{(p+3)/4}
\left (\frac{\theta_{\rmj,\rmw}}{\theta_\rmobs}\right )^{(p+3)/2} \ ,
\label{fb_1}
\eearr
\bearr{}
f_2^b & \equiv & \left.\frac{F_{\nu,\rmn}}{F_{\nu,\rmw}}
\right|_{t=t_{\theta,\rmn}}^{\nu > \nu_c} \nonumber\\
      & = & B^{-(p+2)/4} \left (\frac{E_\rmn}{E_\rmw}\right )^{(p+2)/4}
\left (\frac{\theta_{\rmj,\rmw}}{\theta_\rmobs}\right )^{(p+2)/2} \ ,
\label{fb_2}
\eearr
whereas $f^a$ is given by the inverse of the expression
(eq. [\ref{fa}]) for $t_{\rmdec,\rmw}> t_{\theta,\rmn}$.
In this case the condition $f^b > 1$, which requires $E_\rmn/E_\rmw >
B (\theta_\rmobs/\theta_{\rmj,\rmw})^2$
(see eqs. [\ref{fb_1}] and [\ref{fb_2}]), corresponds to the
narrow component dominating the late afterglow.
If also $f^a > 1$, then the narrow component dominates the entire lightcurve,
with the wide component possibly becoming visible as a bump during the
curve's initial rise. Conversely, when both $f^a$ and $f^b$
are $< 1$, then only the wide component's afterglow emission
would be visible.

If the observer is located outside the solid angle subtended by
the wide outflow component (i.e., $\theta_\rmobs
\gtrsim 1.5\, \theta_{\rmj,\rmw}$), then the flux contributions
from the wide and the narrow components peak at $t_{\theta,\rmw}$ and
$t_{\theta,\rmn}$, respectively. The ratio of these times is
$t_{\theta,\rmw}/t_{\theta,\rmn}
= (E_\rmw/E_\rmn)^{1/3}$, which is independent of $\theta_\rmobs$.
Thus, when $E_\rmw > E_\rmn$ then $t_{\theta,\rmw} > t_{\theta,\rmn}$,
$f_1^a = E_\rmw/E_\rmn$, $f_2^a = (E_\rmw/E_\rmn)^{2/3}$, 
and $f^b = \tilde{f}$. However, if $E_\rmw < E_\rmn$
then $t_{\theta,\rmw} < t_{\theta,\rmn}$,
$f_1^a = E_\rmn/E_\rmw$, $f_2^a = (E_\rmn/E_\rmw)^{2/3}$, 
and $f^b = \tilde{f}^{-1}$. All of these cases have
$f^a > 1$ and $f^b > 1$, which means that the more
energetic outflow component dominates
the lightcurve for $t>\max\{t_{\theta,\rmw},t_{\theta,\rmn}\}$.
\section{Applications}
\label{apply}

Our results have potentially significant implications to the
interpretation of GRBs and XRFs. We consider these two
applications separately, although our discussion makes it clear that
they could be related under a unified picture of these sources.

\subsection{GRB Afterglows and Source Energetics}
\label{apply_GRB}

We choose as fiducial parameters 
$\eta_\rmn = 200,$
$\eta_\rmw=15$,
$\theta_{\rmj,\rmn}=0.05\ {\rm rad}$,
$\theta_{\rmj,\rmw}/\theta_{\rmj,\rmn}=3$, and $p=2.2$.
For this choice the ratio $t_{\rmdec,\rmw}/t_{\rmjet,\rmn}$ of the
deceleration time of the wide component to the jet-break time of the
narrow component (eq. [\ref{t_ratio}]) ranges between 3.0 and 1.4 as
$E_\rmw/E_\rmn$ decreases from 3 to 1/3. The near-coincidence of
these two time scales can have interesting observational
ramifications. In particular, if the flux ratio
$(F_{\nu,\rmw}/F_{\nu,\rmn})_{t=t_{\rmdec,\rmw}}$ is close to
1, then the presence of a break in the
narrow component at $t_{\rmjet,\rmn}$ may be masked by the rise
in the flux from the wide component that occurs as $t_{\rmdec,\rmw}$
is approached. For the adopted parameters, one finds from equations
(\ref{f1})--(\ref{f_end})
that, in fact,
$(F_{\nu,\rmw}/F_{\nu,\rmn})_{t=t_{\rmdec,\rmw}}\gtrsim 1$ so
long as $E_\rmw/E_\rmn$ is $\gtrsim 
%AK: I changed 3 to
2$, but that the flux ratio
becomes $\ll 1$ for low values of the wide-to-narrow injected
energy ratio. For example, in the case described by equation 
(\ref{f1}),
$(F_{\nu,\rmw}/F_{\nu,\rmn})_{t=t_{\rmdec,\rmw}}=
1.9$, $0.3$, and $0.04$ for $E_\rmw/E_\rmn=3$, $1$, and $1/3$, respectively.
For $t>t_{\rmdec,\rmw}>t_{\rmjet,\rmn}$ the flux from the wide
component decreases with time no faster than $t^{-(3p-2)/4}$
($=t^{-1.15}$) or even increases with $t$ (for $t_{\rm
dec,w}<t<\min\{t_{m,\rmw},t_{\rm jet,w}\}$; see
eqs. [\ref{F_wide_start}]--[\ref{F_wide_end}]), whereas the flux
from the narrow component decreases steeply (as $t^{-p}$;
eqs. [\ref{F_narrow_start}]--[\ref{F_narrow_end}]). Thus,
even if $F_{\nu,\rmw}$ is still $< F_{\nu,\rmn}$ at
$t=t_{\rmdec,\rmw}$ (i.e., $f_1<1$), when $E_\rmw>E_\rmn$ it
will become the dominant contributor to the total afterglow flux
soon thereafter. (In the case described by eq.
[\ref{f1}], this will occur for $t/t_{\rmdec,\rmw}>f_1^{-4/(p+2)}$.)
As illustrated by the model lightcurve plotted in Figure
\ref{fig1}$a$, a clear signature of a jet break in the lightcurve
would occur, under these circumstances, only at $t=t_{\rmjet,\rmw}$.

\begin{figure}[t]
  \centering
  \includegraphics*[width=\hsize]{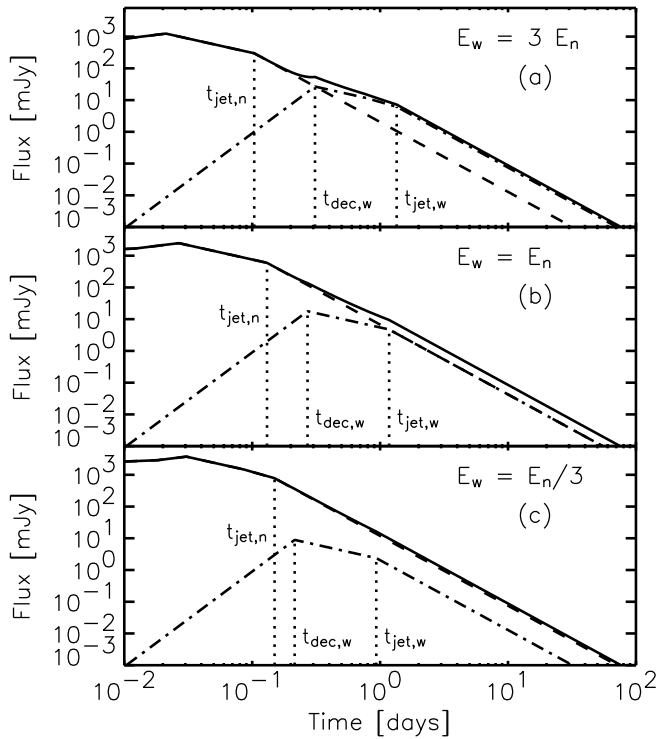}
  \caption
{R-band afterglow lightcurve from a two-component jet. The 
contribution of the narrow component, wide component, and their 
sum is represented by the dashed, dash-dotted, and solid curves, 
respectively. The total outflow energy is assumed to be constant, 
$E_\rmw+E_\rmn = 10^{51}\ {\rm ergs}$. From top to bottom, the 3 panels 
correspond to $E_\rmw/E_\rmn= 3$, 1, and 1/3, respectively. The other
parameters are the same for all panels: 
$\eta_\rmn=200$, 
$\eta_\rmw=15$,
$\theta_{\rmj,\rmw}=0.15$, $\theta_{\rmj,\rmn}=0.05$, $n_0=1$, 
$\epsilon_e=0.1$, $\epsilon_B=0.01$, $p=2.2$, and $D_{\rm L,28}= 1$.}
\label{fig1}
\end{figure}

The possibility that the jet break in the narrow outflow
component remains unobservable could have important consequences
for the inferred energetics of GRBs. Recall that it has been
found \citep[][]{F01,BFK03} that, when the isotropic-equivalent
$\gamma$-ray energies of a sample of GRBs are converted into true
energies by using the beaming factor inferred from the observed jet
break, then the resulting values cluster narrowly about
$E_\gamma \approx 10^{51}\ {\rm ergs}$. On the other hand, the
isotropic-equivalent kinetic energies of GRB outflows, as estimated from
dynamical and spectral modeling of the associated afterglows,
yield --- after being corrected by the jet break-inferred beaming
factor --- a narrow distribution of true kinetic energies (for 2
oppositely directed jets) at the beginning of the afterglow
phase that is centered on $E \approx 10^{51}\ {\rm ergs}$
(\citealt{PK02}; see also \citealt{Y03}). A similar result is
obtained when the afterglow X-ray luminosity is used as a surrogate
for the isotropic-equivalent outflow kinetic energy
\citep[][]{BKF03}. The X-ray emission typically peaks
during the early afterglow, so the kinetic energy estimated on
the basis of the X-ray luminosity (and conventionally evaluated at $t=10\
{\rm hr}$) likely corresponds to that of the {\em narrow} outflow
component. However, the X-ray--based isotropic-equivalent kinetic energy
is typically found to be smaller than the isotropic-equivalent
kinetic energy inferred from the spectral and dynamical modeling
of the overall afterglow (which, for $E_\rmw > E_\rmn$, is dominated by
the contribution of the wide component). In the context of the
two-component jet model this implies that the wide component should
dominate also at early times, which is clearly inconsistent. The X-ray--based
isotropic-equivalent kinetic energy typically also turns out to be smaller
than the isotropic-equivalent $\gamma$-ray energy, which in a
two-component model that associates the $\gamma$-rays with the
narrow core would be difficult to reconcile with the
internal-shock scenario of GRBs (see discussion in the next
paragraph). It is therefore quite possible that the X-ray--based
deduction systematically underestimates the true kinetic energy
in the narrow outflow component.

The approximate equality of the inferred values of $E_\gamma$ and
$E$ has been given several different explanations; here we focus on
its interpretation in the context of the internal-shock scenario for
GRBs \citep{RM94}, which has been successful at accounting for the
observed variability properties of the bursts \citep[e.g.,][]{NP02}.
In this picture, the $\gamma$-ray emission originates in shocks that
form in the collisions of ``shells'' that are injected with variable
energy and/or mass at the origin. It was shown \citep[][]{B00,KS01}
that $E_\gamma/E$ can in principle be $\sim 1$ in this case if the
following conditions are satisfied: (1) the spread between the
minimum and maximum initial Lorentz factors of the shells is large
enough ($\gamma_{\rm i,max}/\gamma_{\rm i,min}\lesssim 10$); (2) the
distribution of initial Lorentz factors is sufficiently nonuniform
(one possibility being that $\log{\gamma_{\rm i}}$, rather than
$\gamma_{\rm i}$, is distributed uniformly); (3) the shells are
approximately of equal mass and their number is large enough
($\gtrsim 30$), and (4) the fraction of the dissipated energy that
is deposited in electrons and then radiated away is sufficiently
high ($\epsilon_e\gtrsim 0.3$), with a similar constraint applying
to the fraction of the radiated energy that is emitted as
gamma-rays. If any of these conditions were violated to a
significant extent then the implied magnitude of $E_\gamma/E$ could
decrease to a value well below 1.

We do not at present have independent information about the
nature of GRB outflows to verify that the
above conditions are indeed satisfied in the majority of
sources, as would be required for consistency between the
internal-shock model and the inferred distributions of
$E_\gamma$ and $E$. It is, however, worth noting that these
constraints can in principle be alleviated if the outflows
correspond to two-component jets with $E_\rmw\gtrsim
E_\rmn$.\footnote{This possibility was originally noted in a
talk at the 2003 GRB meeting in Santa Fe; see \citet{K04}.}
This is because the inference $E_\gamma/E \sim 1$
is based on the assumption that the solid angle used to convert the
isotropic-equivalent energy into a true energy is the same for
the $\gamma$-rays and the afterglow radiation. 
If $E_\rmw\gtrsim E_\rmn$ then most of the afterglow radiation will be emitted
from the wide component and the value of $E$ will be
appropriately inferred from $E_{\rm iso}$ using the opening
half-angle of the wide outflow component,
$\theta_{\rmj,\rmw}$. However, as the prompt high-energy
emission in this picture originates in the narrow outflow component,
the conversion of the isotropic-equivalent $\gamma$-ray energy
into $E_\gamma$ must be done using the opening half-angle of the
narrow component, $\theta_{\rmj,\rmn}$. If, as discussed above,
the jet break in the narrow component is not observationally
discernible and the outflow is mistakenly interpreted as a
single-component jet with an opening half-angle
$\theta_{\rmj,\rmw}$, then 
$E_\gamma$ will be {\em overestimated}
by a factor $\sim (\theta_{\rmj,\rmw}/\theta_{\rmj,\rmn})^2$ ($=9$ for the
fiducial values adopted in this paper). The actual magnitude of
$E_\gamma$ in this case could thus be well below
the value inferred on the basis of a single-component jet
model.

\citet{KP00b} proposed that the ejected
material in GRB outflows exhibits strong angular fluctuations,
forming ``patchy'' shells. In this case the conversion from an
isotropic-equivalent to a true energy also involves a smaller
effective solid angle for the $\gamma$-ray emission than for the
afterglow radiation. This situation could in principle be
distinguished from the two-component outflow scenario discussed
in this paper through some of the specific predictions of each of these
models. For example, \citet{KP00b} argue that a patchy-shell
outflow could exhibit large temporal fluctuations (with a
progressively decreasing amplitude) during the early (minutes to hours)
afterglow, whereas the results derived in this paper indicate
that the afterglow lightcurve produced by a two-component jet with
$E_\rmw\gtrsim E_\rmn$ might temporarily depart from a simple
power-law decay around $t \approx t_{\rmdec,\rmw}$
(over a timescale of hours).\footnote{As discussed below, refreshed
shocks are another likely source of lightcurve variability (on
timescales of hours to days).} 
It is, however, conceivable that the 
ejected shell may be patchy even if the outflow has more than
one component.

As we have shown, the possibility that 
$E_\gamma$ is overestimated in a
two-component outflow can only be realized if $E_\rmw$ exceeds $E_\rmn$.
However, in order for the high-efficiency requirement on the
emission from internal shocks (and the corresponding conditions
listed above) to be relaxed, the ratio $E_\rmw/E_\rmn$ cannot be
much greater than 1. 
Specifically, the kinetic-to-radiative
energy conversion efficiency of the narrow outflow component,
${\cal{E}}_\rmn \equiv
E_\gamma/(E_\gamma + E_\rmn)$, is determined by the ratio
$E_\gamma/E_\rmn$, which is overestimated by the factor
$(E_\rmn/E_\rmw)(\theta_{\rmj,\rmw}/\theta_{\rmj,\rmn})^2 =
E_{\rmiso,\rmn}/E_{\rmiso,\rmw}$ if $E \approx
E_\rmw$.\footnote{If the kinetic energy inferred from the afterglow
fitting in fact corresponds to the sum of the contributions from
the wide and narrow components, then the factor $E_\rmn/E_\rmw$
in the overestimation expression is replaced by
$1/(1+E_\rmw/E_\rmn)$.} Thus, to have any reduction in the required
efficiency, $E_{\rmiso,\rmn}/E_{\rmiso,\rmw}$ must exceed 1.
The above two constraints can be expressed as a double inequality on the
ratio of the component kinetic energies:
\be
1< E_\rmw/E_\rmn<(\theta_{\rmj,\rmw}/\theta_{\rmj,\rmn})^2 \ .
\label{e_con}
\ee 
The condition $E_\rmw/E_\rmn>1$ implies that the wide component
dominates the afterglow emission at late times (see
eqs. [\ref{fmax_tjet}]-[\ref{f2_tilde}]), whereas the requirement
$E_{\rmiso,\rmn}/E_{\rmiso,\rmw}>1$ implies that the narrow
component dominates at early times (see eqs. [\ref{fmax_dec}]-[\ref{f_end}]).

A two-component outflow with $E_\rmw/E_\rmn \gtrsim 1$ arises
naturally in the (initially) neutron-rich, hydromagnetically accelerated
jet scenario (see \S~\ref{introduction}). In contrast, in the
two-component outflow investigated in the context of the collapsar
model, the ratio $E_\rmw/E_\rmn$
is typically $\ll 1$ and hence (see Fig.~\ref{fig1}$c$)
the optical afterglow emission from the wide (cocoon) component
would generally remain undetectable at all times. For the
adopted fiducial parameters and assuming $E_\rmw/E_\rmn=0.1$,
$t_{\rmdec,\rmw}/t_{\rmjet,\rmn}=0.96$ and, for the case
described by equation (\ref{f1}),
$(F_{\nu,\rmw}/F_{\nu,\rmn})_{t=t_{\rmdec,\rmw}}= 5\times10^{-3}$.
Note, however, that the cocoon afterglow
emission might dominate at early times at submillimeter
wavelengths and that there may also be a signature in the early
optical afterglow of the collision between the expanding cocoon
and the decelerating head of the narrow outflow
component \citep{RCR02}.

\begin{figure}[t]
\centering
\includegraphics[width=\hsize]{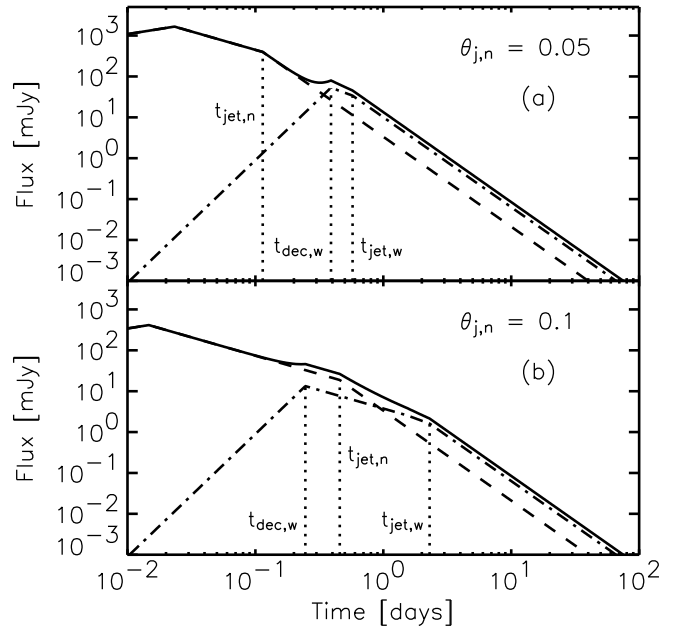}
\caption
{Similar to Fig.~\ref{fig1}, except that $E_\rmw=2\ E_\rmn$ and
$\theta_{\rmj,\rm w} =2\ \theta_{\rmj,\rmn}$. The top and bottom
panels correspond to $\theta_{\rmj,\rmn}=0.05$ and 0.1, respectively.
The other parameters are the same as in Fig.~\ref{fig1}.}
\label{fig2}
\end{figure}

The plots in Figure \ref{fig1}
demonstrate that the optical afterglow lightcurve from a two-component
jet departs from a simple power-law decay when the flux
contributions from the two components become comparable, which
for a jet with $E_\rmw \gtrsim E_\rmn$ typically occurs around
the deceleration time of the wide component. As the wide
component gradually takes over from the narrow component to become the dominant
contributor to the flux, the lightcurve exhibits a concave
``flattening'' (if $t_{\rmdec,\rmw} > t_{\rmjet,\rmn}$; see
Fig.~\ref{fig2}$a$) or a convex ``bump'' (if $t_{\rmdec,\rmw} <
t_{\rmjet,\rmn}$; see Fig.~\ref{fig2}$b$) of duration $\Delta
t \sim t$.  The presence of this feature may
be hard to discern in practice because of insufficiently dense time
coverage or the interference of other factors (the emission from
the reverse shock, ambient density
inhomogeneities, refreshed shocks, etc.) that could cause a similar
behavior. Nonetheless, there are already several potential
candidates for this feature among observed
afterglows. For example, GRB 970508 exhibited a
pronounced brightening between $\sim 1$ and $\sim 2\ {\rm
days}$, after which it followed an approximate power-law decay
\citep[e.g.,][]{P98}. However, between $\sim 0.1$ and $\sim 1\ {\rm
day}$ the lightcurve was constant or slightly declining with
time, a behavior that is not reproduced by our simple
two-component jet model.\footnote{In our picture, the brightening
would be caused by a wide component with $E_\rmw \gtrsim
E_\rmn$ that is observed at $\theta_\rmobs \lesssim \theta_{\rmj,\rmn}$ and
peaks at $t=t_{\rmdec,\rmw}$. In contrast, \citet{PMR98} attributed
the flux rise to a narrow jet seen outside its opening half-angle and
suggested that the flux at earlier times could be produced by a
wide (essentially isotropic) component of lower energy.} A major
brightening event was also recorded in GRB 021004 around $\sim
0.1\ {\rm days}$ \citep[e.g.,][]{F03}. In this case, the
lightcurve assumed a flat form between $\sim 0.02$ and $\sim
0.1\ {\rm days}$ and a power-law decay index of $\sim 1.2$
immediately thereafter \citep[][]{U03}, which is consistent with
the behavior expected in the two-component model (see
Figs. \ref{fig1}a and \ref{fig2}a; if this brightening is
indeed associated with the emergence of the wide-component
emission then the inferred power-law index implies, by
eq. [\ref{F_wide_start}], $p \approx 2.2$ for this component). The
identification of the brightening time with $t_{\rmdec,\rmw}$
suggests that the wide component in this source had a
comparatively high value of $\eta_\rmw$ (see eq. [\ref{t_dec}]).

An alternative interpretation of the early brightening in GRB 021004
was given by \citet{KZ03}, who attributed it to the emission of the
forward shock taking over from that of the reverse shock. It is,
however, worth noting that GRB 021004 exhibited a second, less
pronounced brightening at $t\approx 1\ {\rm day}$ and possibly a third
one at $t\approx 3\ {\rm days}$, and that it has been suggested that
all these events may have a similar physical origin --- either a
variable external density \citep[][]{L02} or energy fluctuations that,
in turn, could arise either from variable injection at the source
(refreshed shocks) or from a patchy angular structure of the
outflow \citep[][]{NPG03}.  Interestingly, a refreshed-shock scenario
is a natural feature of the hydromagnetic, initially neutron-rich jet
model of \citet{VPK03}. In this picture, the decoupled neutrons that
constitute the wide outflow component decay into protons on a distance
scale $\gtrsim R_\beta = 4\times 10^{14}(\eta_\rmw/15)\ {\rm cm}$,
which is likely larger than the scale over which many of the shell
collisions invoked in the internal shock model for GRBs take place.
Shell collisions may well give rise to the $\gamma$-ray emission from
the narrow (proton) outflow component, but since they cannot take
place inside the wide component before the neutrons are converted into
protons, all the fast shells that overtake slower shells at $R
\lesssim R_\beta$ would become arranged in a sequence wherein the
faster shells are in the front and the slower ones lag behind and
remain closer to the origin. Furthermore, the radius at which the
neutron shells are arranged in this way is smaller by a factor of
$\sim(\eta_\rmn/\eta_\rmw)^2$ compared to the radius of the internal
shocks, and is thus $\ll R_\beta$. The neutron shells can therefore
pass through each other with very little interaction between
them while ordering themselves according to their velocities.
After the decay into protons the wide outflow component would start
sweeping up the ambient mass, which would cause it to
decelerate. Under these circumstances, the slower shells that had
been left behind would overtake the decelerated front shell,
leading to a pileup as progressively lower-$\eta$ shells arrive at
correspondingly later times. The wide-component afterglow would then
assume the form of a repeatedly reenergized shock, with the energy
injection occurring quasi-continuously at first and then possibly
tapering off as the slowest shells finally arrive at the front-shock
location. This picture is broadly compatible with the observations of
GRB 021004: the large initial brightening may be interpreted as the
quasi-continuously energized early afterglow emission from the wide
outflow component, and the subsequent rebrightenings may be attributed
either to collisions with late-arriving shells \citep[e.g.,][]{KP00a}
or to a patchy angular structure (for which other aspects of the
afterglow provide independent support; see \citealt{NP03} and
\citealt{NO04}).

Similarly to GRB 021004, GRB 030329 manifested a significant
brightening in its optical lightcurve (at $t \approx 1.5\ {\rm
days}$), followed by several less pronounced rebrightenings (at
$t \approx 2.6$, 3.3, and $5.3\ {\rm days}$, respectively).
\citet{GNP03} and \citet{PNG04} argued that these brightening
episodes can be interpreted in terms of refreshed
shocks in a single-component jet that reenergize the afterglow
emission after the jet-break time (at $\sim 0.5\ {\rm
days}$). An alternative interpretation of the prominent initial
brightening in terms of the emergence of the wide component
in a two-component outflow was given by \citet{B03} and
\citet{S03}. They identified the early break in the optical
lightcurve with $t_{\rmjet,\rmn}$, the initial brightening with
$t_{\rmdec,\rmw}$, and a subsequent break in the radio
lightcurve at $\sim 10\ {\rm days}$ with
$t_{\rmjet,\rmw}$.\footnote{A milder break around $0.25\ {\rm
days}$ can be interpreted as corresponding to the
transition time $t_\rmc$ (eq. [\ref{t_c}]); see \citet{L04}.}
To apply our two-component model to this source, we adopt the
latter interpretation and consider the early afterglow
lightcurve ($t\lesssim 1\ {\rm day}$) as having been dominated
by the narrow outflow component. We incorporate the apparent
presence of refreshed shocks by taking the value of $E_{\rmw}$
at $t_{\rmjet,\rmw}$ as being $\sim 3$ times larger than the
corresponding value at $t_{\rmdec,\rmw}$
\citep[see][]{GNP03}. We adopt $E_{\rmiso,\rmw,52}/n_0 \approx
30$ at $t=t_{\rmjet,\rmw}$ on the basis of the (rather
uncertain) estimates obtained from fitting the sizes of the
radio images of this afterglow at $t=24$ and $83\ {\rm days}$
\citep{T04,GRL05}. Using this value in equation
(\ref{t_jet1}) and $E_{\rmiso,\rmw,52}/n_0\approx 10$ in
equation (\ref{t_dec}), we infer $\theta_{\rmj,\rmw}=0.26$ and
$\eta_\rmw=8.8$, respectively. Adopting $p=2.25$ from the
early-afterglow spectral fit of \citet{W04}, we extend the
narrow-component's flux from $t_{\rmjet,\rmn}$ to
$t_{\rmdec,\rmw}$ and deduce $f_1 =
F_{\nu,\rmw}(t_{\rmdec,\rmw})/F_{\nu,\rmn}(t_{\rmdec,\rmw})\approx
2.3$. Equations (\ref{t_dec}), (\ref{t_jet1}), and (\ref{f1})
then yield $E_{\rmiso,\rmw}/E_{\rmiso,\rmn} = 0.38$
and $\eta_\rmw\theta_{\rmj,\rmn}=0.76$.
We thus infer $\theta_{\rmj,\rmn}\approx 0.086$. As a check on
these deductions, we calculate the flux ratio $\hat{f}_1 =
F_{\nu,\rmw}(t_{\rmjet,\rmw})/F_{\nu,\rmn}(t_{\rmjet,\rmn})$
using equation (\ref{f1_hat}). We obtain $\hat{f}_1 \approx 0.025$,
which agrees well with the observed value of 
$\sim 0.033$ given that $\sim 20\%$ of the observed flux at 
$t_{\rmjet,\rmw}$ appears to
have come from the associated supernova SN 2003dh \citep{B03,L04}.
To estimate the true energy of the two outflow components, we
assume that the measured isotropic-equivalent $\gamma$-ray energy of
this burst ($\simeq 10^{52}\ {\rm ergs}$; \citealt{P03}, \citealt{HS03})
has been produced by the narrow component with a radiative
efficiency $\sim 0.2$, which implies $E_{\rmiso,\rmn,52}\approx 5$
and hence $n_0 \approx 0.19$.
The latter value is consistent with the
density estimates obtained from spectral modeling of the
afterglow \citep[e.g.,][]{B03,W04,GRL05}. In this way we deduce
$E_{\rmw,50} \approx 6.4$ and $E_{\rmn,50}\approx 1.8$.

\begin{figure}[htb]
\centering
\includegraphics*[width=\hsize]{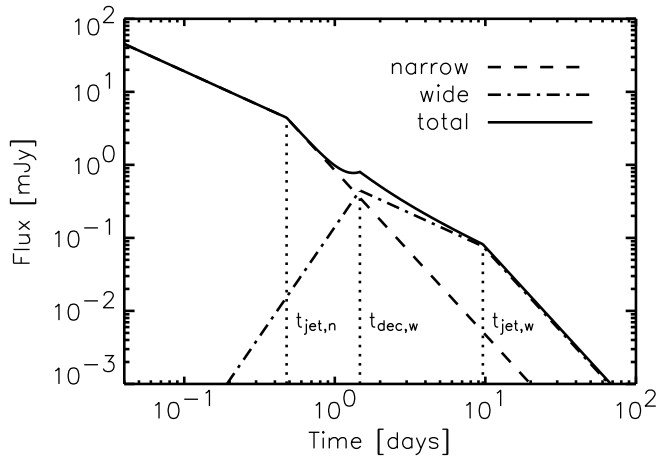}
\caption
{R-band lightcurve from a two-component model with parameters
appropriate to GRB 030329 ($D_{\rm L}=0.8\ {\rm Gpc}$):
$\eta_\rmn=200$, $\eta_\rmw=8.8$, $E_\rmw=6.4\times 10^{50}\ {\rm ergs}$, 
$E_\rmn=1.8\times 10^{50}\ {\rm ergs}$, $\theta_{\rmj,\rmw}=0.26$,
$\theta_{\rmj,\rmn}=0.086$, $n_0=0.19\ {\rm cm^{-3}}$,
$\epsilon_e=0.01$, $\epsilon_B=0.008$, and $p=2.25$.}
\label{fig3}
\end{figure}

An illustrative two-component model lightcurve based
on the above estimates is shown in Figure \ref{fig3}.
The values of the parameters $\epsilon_e$ and $\epsilon_B$
in this fit were chosen somewhat arbitrarily to approximate the
measured flux level (which is mostly sensitive to $\epsilon_e$).
The value of $\eta_\rmn$ cannot be inferred from the available
observations and was simply chosen to be $\gg \eta_\rmw$.
Although this simple model can account for the presence of the
pronounced bump at $t \approx 1.5\ {\rm days}$, the model
lightcurve (which in general cannot rise faster than $\propto
t^3$; see eqs. [\ref{F_wide_start}]--[\ref{F_wide_end}]) cannot
readily reproduce the sharpness of the observed flux increase at
that time. The steep rise might, however, be explained by the
refreshed-shock model \citep[e.g.,][]{GNP03}. This possibility
is consistent with observational evidence that the ``shock
refreshing'' process in this source was already under way at
$t=t_{\rmdec,\rmw}$ \citep{W04,L04}. As noted above in connection
with GRB 021004, the occurrence of such shocks around
$t_{\rmjet,\rmw}$ can be naturally expected in the
hydromagnetic, initially neutron-rich outflow scenario. It is
worth pointing out, however, that other potential problems with the
two-component interpretation of the GRB 030329 data still remain
to be addressed. In particular \citep[see][]{PNG04}, the strong
radio signature from a reverse shock that is expected in this
picture at $t\approx t_{\rmd,\rmw}$ has not been detected.

The afterglows of GRB 021004 and GRB 030329 were distinguished by
the fact that their monitoring started early on and was conducted
with particularly high time resolution and precision. Future
observations will determine whether the departure from a smooth
power-law behavior exhibited by the lightcurves from these two
sources is a common feature of GRB afterglows. If this turns out
to be the case, then the possibility that this behavior is
associated with the decoupling of a neutron component in a
hydromagnetically driven jet would merit a closer examination.

\subsection{XRF Afterglows and Source Energetics}
\label{apply_XRF}

X-ray flash (XRF) sources \citep[e.g.,][]{H03,K03} are high-energy
transients that strongly resemble GRBs except that their peak energies
fall in the X-ray, rather than the $\gamma$-ray, spectral regime. One
attractive interpretation of these sources is that they represent
essentially uniform GRB jets that are observed outside the jet
half-opening angle, $\theta_\rmobs > \theta_\rmj$
\citep[e.g.,][]{Y02,Y04}. In this picture, the larger viewing angle
results in a smaller Doppler factor and hence a lower apparent peak
frequency than in GRBs, which correspond to
$\theta_\rmobs < \theta_\rmj$. The association with GRBs has received
support from the detection of afterglow emission in several XRF
sources. In particular, \citet{S04} carried out the first
spectroscopic observations of an XRF and form modeling of the radio
afterglow of XRF 020903 inferred that its total kinetic energy is
comparable to that typically deduced in GRB sources.  The relatively
low isotropic equivalent energy of the prompt-emission in XRF 020903,
$E_{\rm X,iso}\approx 1.1\times 10^{49}\;{\rm ergs}$, is
$\sim 2$ orders of magnitude smaller than the narrowly
clustered values for the true energy output in gamma-rays deduced for
GRBs, $E_\gamma\approx 10^{51}\;{\rm ergs}$
\citep{F01,BFK03}, which is itself $\sim 1-3$ orders of magnitude
smaller than the isotropic equivalent gamma-ray energy output in GRBs,
$E_{\gamma,{\rm iso}}\sim 10^{52}-10^{54}\;$ergs.
This is consistent with the expected reduction in the measured fluence
for off-axis observers as a result of the decrease in the Doppler
factor.\footnote{The fluence decreases as the third power of the
Doppler factor \citep{G02}, one power from the reduced energy of
each photon and a power of two from the relativistic beaming of the
photons (aberration of light) away from our line of sight.}

In the context of a two-component outflow model with $E_\rmw
\gtrsim E_\rmn$, the above unified GRB/XRF picture leads to the
identification of XRFs with GRB outflows that are observed at
$\theta_\rmobs>\theta_{\rmj,\rmn}$ but likely still within the
opening half-angle of the wide component. Using our fiducial
model parameters and assuming $\theta_\rmobs =
2\, \theta_{\rmj,\rmn}$, we find that
$t_{\rmdec,\rmw}/t_{\theta,\rmn}= 0.65$ and 0.75
(eq. [\ref{t_ratio1}]) and that
$[F_{\nu,\rmn}/F_{\nu,\rmw}](t=t_{\theta,\rmn})= 0.43$ and $0.28$
(eq. [\ref{fb_1}]) for $E_\rmw/E_\rmn = 2$ and 3, respectively.
Thus, even though the wide component would dominate the overall
afterglow emission, the narrow component might give rise to a
bump in the optical lightcurve around $t_{\theta,\rmn}$. In fact,
a bump in the afterglow lightcurve of XRF 030723
was interpreted along these lines by \citet{H04}, who employed a
two-component outflow model with $\theta_{\rmj,\rmw} \approx 3\,
\theta_{\rmj,\rmn}$ and $E_\rmn \approx 3\, E_\rmw$. To account
for the relatively late occurrence of this bump (between $\sim 11$ and
$\sim 14\ {\rm days}$), a rather large observation angle was
adopted in this fit ($\theta_{\rmobs} = 0.37\, {\rm rad}\approx 4\,
\theta_{\rmj,\rmn}$; see eq. [\ref{t_theta}]). However, the
rebrightening of XRF 030723 was found to be accompanied by a
significant spectral reddening \citep{F04}, which is not
naturally explained in the two-component model but could possibly
be associated with a supernova. Further tests of this aspect of
the two-component model would therefore need to await the
detection of additional examples of bumps in XRF afterglow
lightcurves by future observations.

An alternative interpretation of XRFs has been proposed in the
context of the $E_\rmn > E_\rmw$ two-component collapsar outflow
model \citep[e.g.,][]{ZWH04}. In this picture, the transient
X-ray emission is attributed to external shocks driven by the
wide (cocoon) component. Although the detailed implications of
this proposal have not yet been fully worked out, this scenario could
conceivably account for the large inferred ratio of the
transient radiation energy and the afterglow kinetic energy in XRF 020903
by associating the afterglow emission with the external shock
of the more energetic narrow (jet-core) component. However,
since the afterglow emission from the narrow component only
becomes observable for $t>t_{\theta,\rmn}$, it should exhibit a
rapid decline with time, which may not be consistent with the
data from XRF 020903 (where \citealt{S04} still detected
radio afterglow emission after $\sim 200\ {\rm days}$).

\section{Conclusions}
\label{conclude}

In this paper we study the optical afterglow lightcurves
produced by GRB sources that have two distinct outflow components. The
possibility that GRB jets have a narrow/fast core and a
wider/slower outer component has been indicated by observations
of both the $\gamma$-ray and the afterglow emission (including
the afterglows of GRB 970508, GRB 991216, and GRB 030329) and
independently by theoretical considerations. Since we are
interested in the distinct afterglow signatures of the two
components, we focus on models in which the Lorentz factors of
both outflow components are initially $\gg 1$. We choose the
hydromagnetically driven, initially neutron-rich jet model of
\citet{VPK03} and the collapsar jet-breakout model of
\citet{ZWH04} as being representative of scenarios in which the
dual nature of the outflow reflects initial conditions and
propagation effects, respectively. In both of these
models the characteristic initial Lorentz factor and opening
half-angles are $\eta_\rmn \gtrsim 10^2$ and $\theta_{\rmj,\rmn}
\sim 0.05$ for the narrow component and $\eta_\rmw \sim 10$,
$\theta_{\rmj,\rmw} \lesssim 3\, \theta_{\rmj,\rmn}$ for the
wide one, and the $\gamma$-ray emission originates in the narrow
component. They are distinguished, however, by the ratio of
the kinetic energy injected into the two components:
$E_\rmw/E_\rmn \sim 0.1$ for the collapsar model and $\gtrsim 2$
for the neutron-rich hydromagnetic model (with $E_\rmw + E_\rmn$
inferred to be $\sim 10^{51}\ {\rm ergs}$).

Using a simple synchrotron emission model, we calculate the
afterglow emission produced by the shocks that the two
components drive into the ambient medium. We derive useful algebraic
expressions for the component flux ratios at the main
transition times in the light curve
(in particular, the wide component's
deceleration time $t_{\rmdec,\rmw}$ and the jet-break times
$t_{\rmjet,\rmw}$ and $t_{\rmjet,\rmn}$) for the cases where the
observation angle $\theta_\rmobs$ satisfies
$\theta_\rmobs < 1.5\,\theta_{\rmj,\rmn}$ and
$1.5\,\theta_{\rmj,\rmn}<\theta_\rmobs<1.5\,\theta_{\rmj,\rmw}$,
respectively (where in the latter case $t_{\theta,\rmn}$ is the
relevant break time for the narrow component). We study the behavior
of the optical lightcurves for different values of $E_\rmw/E_\rmn$ and
find that, for the adopted characteristic parameters, the
contribution of the narrow component dominates at all times if
this ratio is $\ll 1$ (as in the collapsar jet-breakout model),
but that the contribution of the wide component becomes dominant
for $t \gtrsim t_{\rmdec,\rmw}$ if $E_\rmw \gtrsim 2\, E_\rmn$
(as in the neutron-rich hydromagnetic model). The emergence of
the wide component may be related to the pronounced brightening
detected in the lightcurves of several afterglows $\sim 0.1-1\
{\rm days}$ after the GRB (see Fig.~\ref{fig3}).

For typical parameter values $t_{\rmdec,\rmw}$ is found to be
comparable to $t_{\rmjet,\rmn}$. It follows that, if $E_\rmw >
E_\rmn$, then the steepening of the narrow component's lightcurve
at $t \gtrsim t_{\rmjet,\rmn}$ could be masked by the emergence (and
subsequent dominance) of the wide component (see Fig.~\ref{fig1}).
Under these circumstances, the only clearly discernible jet break in
the optical lightcurve would occur at $t_{\rmjet,\rmw}$. We
suggest that this may have led to an overestimate of the emitted
$\gamma$-ray energy in many GRBs because the wide component's
opening half-angle $\theta_{\rmj,\rmw}$ --- rather than the
narrow component's angle $\theta_{\rmj,\rmn}$ --- was used in
converting the measured $E_{\gamma,\rmiso}$ into the true energy
$E_\gamma$. This, in turn, would have led to an overestimate [by
a factor $\sim (E_\rmn/E_\rmw)(\theta_{\rmj,\rmw}/\theta_{\rmj,\rmn})^2
=E_{\rmiso,\rmn}/E_{\rmiso,\rmw}$]
of the ratio $E_\gamma/E_\rmn$ that determines the
kinetic-to-radiative energy conversion efficiency of the
outflow. Factoring in this overestimate 
[which can be done when the component kinetic energies satisfy $1<
E_\rmw/E_\rmn< (\theta_{\rmj,\rmw}/\theta_{\rmj,\rmn})^2$]
would alleviate the need to account for conversion efficiencies
$O(1)$ in internal shock models of GRBs. Dense
monitoring of the afterglow lightcurve during the time interval
($\sim 0.1-1\ {\rm days}$) that encompasses $t_{\rmjet,\rmn}$
and $t_{\rmdec,\rmw}$ could provide a test of this
suggestion. If $t_{\rmjet,\rmn} \lesssim t_{\rmdec,\rmw}$ then
the lightcurve should exhibit a convex bump during this time
interval, whereas if this inequality is reversed a concave
flattening would be expected (see Fig.~\ref{fig2}).
The above considerations also apply to jets observed at
$\theta_\rmobs\gtrsim 1.5\, \theta_{\rmj,\rmn}$, which would be
perceived as X-ray flash sources. For $E_\rmw > E_\rmn$, the
afterglow emission from such sources would be dominated by the
wide outflow component, although the narrow component might give
rise to a bump in the lightcurve at $t \approx t_{\theta,\rmn}$.

The well-monitored afterglow lightcurves of GRB 021004 and
GRB 030329 exhibited a significant early brightening that was
followed by several less pronounced rebrightenings on a time scale
of days. These episodes can be satisfactorily interpreted as
refreshed shocks. We point out that the initially neutron-rich
hydromagnetic jet model, in which the decoupled protons and
neutrons give rise to a narrow/fast and wide/slow outflow
components, respectively, could naturally account for the
appearance of refreshed shocks following (or even coincident with)
the wide component's emergence in the afterglow lightcurve.
Future high-quality observations should be able to determine
whether the nonsteady behavior found in these two objects is a
common trait of GRB sources.

\acknowledgments 

We thank Andrei Beloborodov and Nektarios Vlahakis
for helpful conversations and Yizhong Fan, Xuefeng Wu, and Bing Zhang for
useful correspondence.
We are also grateful to the referee for his careful reading of the
manuscript and constructive comments.
This research was supported in part by NASA
Astrophysics Theory Program grant NAG5-12635. FP acknowledges support
from the Department of Energy under grant B341495 to the Center for
Astrophysical Thermonuclear Flashes at the University of Chicago.
JG acknowledges support by the U.S. Department of
Energy under contract DE-AC03-76SF00515 to Stanford University.

\appendix

\section{R-Band Fluxes From The Two Outflow Components}
\label{afterglow}
For the narrow/fast jet component, typically $t_{\rm dec,n}$ is
smaller than all the other transition times and $t_{0,\rm n}<t_{\rmjet,n}$,
so there are four interesting regimes of observed frequencies: $
\nu > \nu_{0,\rmn}; \quad \nu_{\rmc,\rmn}(t_{\rmjet,\rmn}) <
\nu < \nu_{0,\rmn}; \quad \nu_{\rmm,\rmn}(t_{\rmjet,\rmn}) <
\nu < \nu_{\rmc,\rmn} (t_{\rmjet,\rmn}); \quad \nu  <
\nu_{\rmm,\rmn}(t_{\rmjet,\rmn}) $. The corresponding fluxes are
\bearr{}
\frac{F_{\nu,\rmn}}{F_{\nu,{\rmmax,n}}}
& = &
        \left \{\begin{array}{ll}
        (t_{\rmdec,\rmn}/t_{\rmc,\rmn})^{1/6}
          (t/t_{\rmdec,\rmn})^{11/3} & t < t_{\rmdec,\rmn}\ ,\\
        (t/t_{\rmc,\rmn})^{1/6} &
          t_{\rmdec,\rmn} < t < t_{\rmc,\rmn}\ ,\\
        (t/t_{\rmc,\rmn})^{-1/4}  &
          t_{\rmc,\rmn} < t < t_{\rmm,\rmn}\ ,\\
        (t_{m,\rmn}/t_{\rmc,\rmn})^{-1/4} (t/t_{m,\rmn})^{-(3p-2)/4} &
          t_{\rmm,\rmn} < t < t_{\rmjet,\rmn}\ ,\\
        (t_{m,\rmn}/t_{\rmc,\rmn})^{-1/4}
          (t_{\rmjet,\rmn}/t_{m,\rmn})^{-(3p-2)/4}
        (t/t_{\rmjet,\rmn})^{-p}\ \ & t > t_{\rmjet,\rmn}\ ,
                \end{array}
        \right.
\label{F_narrow_start}
\eearr{}
\bearr{}
\frac{F_{\nu,\rmn}}{F_{\nu,{\rm max,n}}} & = &
        \left \{\begin{array}{ll}
        (t_{0,\rmn}/t_{\rmm,\rmn})^{1/2}(t_{\rmdec,\rmn}/t_{0,\rmn})^{1/6}
          (t/t_{\rmdec,\rmn})^{11/3} & t < t_{\rmdec,\rmn}\ ,\\
        (t_{0,\rmn}/t_{\rmm,\rmn})^{1/2}(t/t_{0,\rmn})^{1/6} &
          t_{\rmdec,\rmn} < t < t_{0,\rmn}\ ,\\
        (t/t_{\rmm,\rmn})^{1/2} & t_{0,\rmn} < t < t_{\rmm,\rmn}\ ,\\
        (t/t_{\rmm,\rmn})^{-3(p-1)/4} &
          t_{\rmm,\rmn} < t < t_{\rmc,\rmn}\ ,\\
        (t_{\rmc,\rmn}/t_{\rmm,\rmn})^{-3(p-1)/4}
          (t/t_{\rmc,\rmn})^{-(3p-2)/4}
          & t_{\rmc,\rmn} < t < t_{\rmjet,\rmn}\ ,\\
        (t_{\rmc,\rmn}/t_{\rmm,\rmn})^{-3(p-1)/4}(t_{\rmjet,\rmn}/
          t_{\rmc,\rmn})^{-(3p-2)/4}(t/t_{\rmjet,\rmn})^{-p}\ \ &
          t > t_{\rmjet,\rmn}\ ,
                \end{array}
        \right.
\label{F_narrow_2}
\eearr{}
\bearr{}
\frac{F_{\nu,\rmn}}{F_{\nu,{\rm max,n}}} & = &
        \left \{\begin{array}{ll}
        (t_{0,\rmn}/t_{\rmm,\rmn})^{1/2}(t_{\rmdec,\rmn}/t_{0,\rmn})^{1/6}
          (t/t_{\rmdec,\rmn})^{11/3}\ \ & t < t_{\rmdec,\rmn}\ ,\\
        (t_{0,\rmn}/t_{\rmm,\rmn})^{1/2}(t/t_{0,\rmn})^{1/6} &
          t_{\rmdec,\rmn} < t < t_{0,\rmn}\ ,\\
        (t/t_{\rmm,\rmn})^{1/2} & t_{0,\rmn} < t < t_{\rmm,\rmn}\ ,\\
        (t/t_{\rmm,\rmn})^{-3(p-1)/4} &
          t_{\rmm,\rmn} < t < t_{\rmjet,\rmn}\ ,\\
        (t_{\rmjet,\rmn}/t_{\rmm,\rmn})^{-3(p-1)/4}
          (t/t_{\rmjet,\rmn})^{-p} & t > t_{\rmjet,\rmn}\ ,
                \end{array}
        \right.
\label{F_narrow_3}
\eearr{}
\bearr{}
\label{F_n,last}
\frac{F_{\nu,\rmn}}{F_{\nu,\rmn}(t_{\rmjet,\rmn})} & = &
        \left \{\begin{array}{ll}
        (t_{0,\rmn}/t_{\rmjet,\rmn})^{1/2}(t_{\rmdec,\rmn}/t_{0,\rmn})^{1/6}
          (t/t_{\rmdec,\rmn})^{11/3}\ \ & t < t_{\rmdec,\rmn}\ ,\\
        (t_{0,\rmn}/t_{\rmjet,\rmn})^{1/2}(t/t_{0,\rmn})^{1/6} &
          t_{\rmdec,\rmn} < t < t_{0,\rmn}\ ,\\
        (t/t_{\rmjet,\rmn})^{1/2} & t_{0,\rmn} < t < t_{\rmjet,\rmn}\ ,\\
        (t/t_{\rmjet,\rmn})^{-1/3}& t_{\rmjet,\rmn} <t < t_{\rmm,\rmn}\ ,\\
        (t_{\rmm,\rmn}/t_{\rmjet,\rmn})^{-1/3}(t/t_{\rmm,\rmn})^{-p} &
          t > t_{\rmm,\rmn}\ .
                \end{array}
        \right.
\label{F_narrow_end}
\eearr{}
It is worth noting at this point that the simple power-law
scalings we employ in this paper may not always give an
accurate representation of the actual behavior of a real
outflow. Possible discrepancies have, in fact, been indicated
by the hydrodynamical simulations reported in \citet{G01}. For
example, the scaling $F_\nu \propto t^{-1/3}$ given in
the next-to-last line of equation (\ref{F_n,last}) appears to differ
from the behavior exhibited in Figure 2 of that reference, where
the flux at low frequencies is shown to continue rising well past
$t_{\rmjet}$. However, the expressions that are most relevant to
the behavior of the optical afterglow seem to be consistent with the
results of the numerical simulations.

For the wide/slow jet component, the cooling frequency $\nu_c$
is always larger than the characteristic frequency $\nu_m$ for
typical parameters, so it is always slow cooling. This can be verified
in the following analysis. Since the ratio $\nu_c/\nu_m$ decreases with
time before $t_{\rmdec}$ and increases with time after $t_{\rmdec}$,
it reaches its
minimum value at $t_{\rmdec}$, and the condition for always being in slow
cooling is
$(\nu_c/\nu_m)|_{t_{\rmdec,\rmw}}> 1$. Combining equations
(\ref{nu_m}) and (\ref{nu_c}), we get
\be
(\nu_c/\nu_m)|_{t_{\rmdec,\rmw}} = 3.8\,  g^{-2} E_{\rm iso,w,52}^{-2/3}
 \epsilon_{B,-1}^{-2} \epsilon_{e,-1}^{-2} n_0^{-4/3}
 (\eta_\rmw/10)^{-8/3}\ .
\label{nuc_num}
\ee
For the typical parameter ranges described at the end of this
Appendix, $(\nu_c/\nu_m)|_{t_{\rmdec,\rmw}} \sim 5\times 10^{-5} -
2\times 10^9$, which shows that $(\nu_c/\nu_m)|_{t_{\rmdec,\rmw}} > 1$ for
most parameter values. This inequality is violated only when all
the relevant parameters are close to their maximum values, which
is not a typical situation. If, in addition,
$\nu_{\rmc,\rmw}(t_{\rmjet,\rmw}) >
\nu_{\rmm,\rmw}(t_{\rmdec,\rmw})$ is also satisfied (which is
again true for most parameter values), then there are five
interesting frequency regimes: $\nu  > \nu_{\rmc,\rmw}(t_{\rmdec,\rmw})$,
$\nu_{\rmc,\rmw}(t_{\rmjet,\rmw})<\nu <\nu_{\rmc,\rmw}(t_{\rmdec,\rmw})$,
$\nu_{\rmm,\rmw}(t_{\rmdec,\rmw})<\nu <\nu_{\rmc,\rmw}(t_{\rmjet,\rmw})$,
$\nu_{\rmm,\rmw}(t_{\rmjet,\rmw})<\nu <\nu_{\rmm,\rmw}(t_{\rmdec,\rmw})$,
and $\nu  < \nu_{\rmm,\rmw}(t_{\rmjet,\rmw})$. The corresponding fluxes are
\bearr{}
\frac{F_{\nu,\rmw}}{F_{\nu,\rmw}(t_{\rmdec,\rmw})}  & = &
      \left \{\begin{array}{ll}
      (t_{\rmc,\rmw}/t_{\rmdec,\rmw})^2 (t/t_{\rmc,\rmw})^3
        & t < t_{\rmc,\rmw}\ ,\\
      (t/t_{\rmdec,\rmw})^2 & t_{\rmc,\rmw} < t < t_{\rmdec,\rmw}\ ,\\
      (t/t_{\rmdec,\rmw})^{-(3p-2)/4}
        & t_{\rmdec,\rmw} < t < t_{\rmjet,\rmw}\ ,\\
      (t_{\rmjet,\rmw}/t_{\rmdec,\rmw})^{-(3p-2)/4}(t/t_{\rmjet,\rmw})^{-p}
      \ \ & t > t_{\rmjet,\rmw}\ ,\\
       \end{array}
       \right.
\label{F_wide_start}
\eearr{}
\bearr{} \frac{F_{\nu,\rmw}}{F_{\nu,\rmw}(t_{\rmdec,\rmw})}
& = &
      \left \{\begin{array}{ll}
      (t/t_{\rmdec,\rmw})^3 & t < t_{\rmdec,\rmw}\ ,\\
      (t/t_{\rmdec,\rmw})^{-3(p-1)/4} & t_{\rmdec,\rmw}<t< t_{\rmc,\rmw}\ ,\\
      (t_{\rmc,\rmw}/t_{\rmdec,\rmw})^{-3(p-1)/4}(t/t_{\rmc,\rmw})^{-(3p-2)/4}
        & t_{\rmc,\rmw} < t < t_{\rmjet,\rmw}\ ,\\
      (t_{\rmc,\rmw}/t_{\rmdec,\rmw})^{-3(p-1)/4}
        (t_{\rmjet,\rmw}/t_{\rmc,\rmw})^{-(3p-2)/4}(t/t_{\rmjet,\rmw})^{-p}
        \ \ & t > t_{\rmjet,\rmw}\ ,\\
       \end{array}
       \right.
\label{F_wide_2}
\eearr{}
\bearr{}
\frac{F_{\nu,\rmw}}{F_{\nu,\rmw}(t_{\rmdec,\rmw})}
& = &
      \left \{\begin{array}{ll}
      (t/t_{\rmdec,\rmw})^3 & t < t_{\rmdec,\rmw}\ ,\\
      (t/t_{\rmdec,\rmw})^{-3(p-1)/4} &
        t_{\rmdec,\rmw} <t < t_{\rmjet,\rmw}\ ,\\
      (t_{\rmjet,\rmw}/t_{\rmdec,\rmw})^{-3(p-1)/4}
        (t/t_{\rmjet,\rmw})^{-p}\ \ & t > t_{\rmjet,\rmw}\ ,\\
       \end{array}
       \right.
\label{F_wide_3}
\eearr{}
\bearr{}
\frac{F_{\nu,\rmw}}{F_{\nu,{\rm max,w}}} & = &
         \left \{\begin{array}{ll}
      (t_{\rmdec,\rmw}/t_{\rmm,\rmw})^{1/2}
       (t/t_{\rmdec,\rmw})^3 & t < t_{\rmdec,\rmw}\ ,\\
      (t/t_{\rmm,\rmw})^{1/2} & t_{\rmdec,\rmw} < t < t_{\rmm,\rmw}\ ,\\
      (t/t_{\rmm,\rmw})^{-3(p-1)/4} & t_{\rmm,\rmw}<t< t_{\rmjet,\rmw}\ ,\\
      (t_{\rmjet,\rmw}/t_\rmm)^{-3(p-1)/4}(t/t_{\rmjet,\rmw})^{-p}
        & t > t_{\rmjet,\rmw}\ ,\\
                \end{array}
        \right.
\label{F_wide_4}
\eearr{}
\bearr{}
\frac{F_{\nu,\rmw}}{F_{\nu,\rmw}(t_{\rmjet,w})}
& = &
         \left \{\begin{array}{ll}
      (t_{\rmdec,\rmw}/t_{\rmjet,\rmw})^{1/2}
       (t/t_{\rmdec,\rmw})^3\ \ & t < t_{\rmdec,\rmw}\ ,\\
      (t/t_{\rmjet,w})^{1/2} & t_{\rmdec,\rmw} < t < t_{\rmjet,\rmw}\ ,\\
      (t/t_{\rmjet,\rmw})^{-1/3} & t_{\rmjet,\rmw} < t < t_{\rmm,\rmw}\ ,\\
      (t_{\rmm,\rmw}/t_\rmjet)^{-1/3}(t/t_{\rmm,\rmw})^{-p}
        & t > t_{\rmm,\rmw}.\\
                \end{array}
        \right.
\label{F_wide_5}
\eearr{}

If, on the other hand, $\nu_{\rmc,\rmw}(t_{\rmjet,\rmw}) <
\nu_{\rmm,\rmw}(t_{\rmdec,\rmw})$, then there is an additional
possible shape for the lightcurve when $
\nu_{\rmc,\rmw}(t_{\rmjet,\rmw}) < \nu  <
\nu_{\rmm,\rmw}(t_{\rmdec,\rmw})$:
\bearr{}
\frac{F_{\nu,\rmw}}{F_{\nu,{\rm max,w}}} & = &
         \left \{\begin{array}{ll}
      (t_{\rmdec,\rmw}/t_{\rmm,\rmw})^{1/2}
       (t/t_{\rmdec,\rmw})^3 & t < t_{\rmdec,\rmw}\ ,\\
      (t/t_{\rmm,\rmw})^{1/2} & t_{\rmdec,\rmw} < t < t_{\rmm,\rmw}\ ,\\
      (t/t_{\rmm,\rmw})^{-3(p-1)/4}  & t_{\rmm,\rmw}<t< t_{\rmc,\rmw}\ ,\\
      (t_{\rmc,\rmw}/t_\rmm)^{-3(p-1)/4}(t/t_{\rmc,\rmw})^{-(3p-2)/4}
        & t_{\rmc,\rmw} < t < t_{\rmjet,\rmw}\ , \\
      (t_{\rmc,\rmw}/t_\rmm)^{-3(p-1)/4}
       (t_{\rmjet,\rmw}/t_{\rmc,\rmw})^{-(3p-2)/4}(t/t_{\rmjet,\rmw})^{-p}
        \ \ & t > t_{\rmjet,\rmw}\ .\\
                \end{array}
        \right.
\label{F_wide_end}
\eearr{}

In this work we are interested in the R-band spectral regime, so
we compare the characteristic frequencies at the main
spectral transition times with a typical optical frequency
$\nu_{\rm R}$. In particular, at $t_{\rmdec,\rmw}$, \bearr{}
\nu_{\rmc,\rmw}(t_{\rmdec,\rmw}) & = & 5.2 \times10^{13}
\:E_{\rmiso,\rmw,52}^{-2/3}
\epsilon_{B,-1}^{-3/2}n_0^{-5/6}(\eta_\rmw/10)^{4/3}\;{\rm Hz}\ ,\\
\label{nucw_tdecw}
\nu_{\rmc,\rmn}(t_{\rmdec,\rmw}) & = & 
  \left \{\begin{array}{ll}
     \nu_{\rmc,\rmw}(t_{\rmdec,\rmw})
      \left (\frac{E_{\rmiso,\rmw}}{E_{\rmiso,\rmn}}\right )^{1/2} & 
     ~~t_{\rmdec,\rmw} < t_{\rmjet,\rmn} \ , \\ 
     \nu_{\rmc,\rmn}(t_{\rmjet,\rmn}) &
     ~~t_{\rmdec,\rmw} > t_{\rmjet,\rmn} \ ,
          \end{array}
  \right. \\
\label{nucn_tdecw}
\nu_{\rmm,\rmw}(t_{\rmdec,\rmw}) & = & 1.4 \times10^{13}\:g^2\epsilon_{e,-1}^2
\epsilon_{B,-1}^{1/2}n_0^{1/2}(\eta_\rmw/10)^4\;{\rm Hz}\ ,\\
\label{numw_tdecw}
\nu_{\rmm,\rmn}(t_{\rmdec,\rmw}) & = & 
  \left \{\begin{array}{ll}
     \nu_{\rmm,\rmw}(t_{\rmdec,\rmw})
       \left (\frac{E_{\rmiso,\rmw}}{E_{\rmiso,\rmn}}\right )^{-1/2} & 
     ~~t_{\rmdec,\rmw} < t_{\rmjet,\rmn} \ , \\
     \nu_{\rmm,\rmn}(t_{\rmjet,\rmn}) 
       \left (\frac{E_{\rmiso,\rmw}}{E_{\rmiso,\rmn}}\right )^{-2/3}
       \left (\eta_\rmw \theta_{\rmj,\rmn}\right )^{16/3}   &
     ~~t_{\rmdec,\rmw} > t_{\rmjet,\rmn} \ .
          \end{array}
  \right. 
\label{numn_tdecw}
\eearr{}
We also have \bearr{}
\nu_\rmc(t_\rmjet) & = & 1.6 \times 10^{14}\:E_{\rmiso,52}^{-2/3}
\epsilon_{B,-1}^{-3/2}n_0^{-5/6}\theta_{\rmj,-1}^{-4/3}\;{\rm Hz}\ ,\\
\nu_\rmm(t_\rmjet) & = & 1.4 \times10^{13}\:g^2\epsilon_{e,-1}^2
\epsilon_{B,-1}^{1/2}n_0^{1/2}\theta_{\rmj,-1}^{-4}\;{\rm Hz}\ .
\eearr

We adopt as typical parameter ranges $p\sim 1.5-3$, $n_0=0.3-30$,
$\epsilon_e=0.005-0.3$, $\epsilon_B=0.001-0.1$,
$\theta_{\rmj,\rmn}\sim 0.05-0.1$,
$\theta_{\rmj,\rmw}\sim 0.1-0.3$, $E_\rmn \sim E_\rmw \sim
10^{50}-10^{51}\ {\rm ergs}$ \citep{PK02}, $\eta_\rmn\sim 10^2$, and
$\eta_\rmw \sim 10$.
For this set of parameters we find that the following
inequalities are always obeyed:
$\nu_{\rmm,\rmn}(t_{\rmdec,\rmn}) > \nu_{\rm R} >
\nu_{\rmm,\rmn}(t_{\rmjet,n})$,
$\nu_{\rm R} > \nu_{\rmm,\rmn}(t_{\rmdec,\rmw})$, and
$\nu_{\rm R} > \nu_{\rmm,\rmw}(t_{\rmdec,\rmw})$. The
observation frequency
$\nu_{\rm R}$ can be larger or smaller than $\nu_\rmc$ for both components at
$t_{\rmjet,\rmn}$ and $t_{\rmdec,\rmw}$. However, for 
the parameter combinations employed in the plots shown in this paper,
$\nu_{\rmc,\rmn}(t_{\rmjet,\rmn}) > \nu_{\rm R} >
\nu_{\rmm,\rmn}(t_{\rmjet,\rmn})$ 
and
$\nu_{\rmc,\rmw}(t_{\rmdec,\rmw}) > \nu_{\rm R} > 
\nu_{\rmm,\rmw}(t_{\rmdec,\rmw})$.
Since $\nu_m$ and $\nu_m/\nu_c$ decrease with time (see
eqs. [\ref{nu_m2}] and [\ref{nu_c2}]), the inequalities 
$\nu_\rmc > \nu_\rmm$  and $\nu_{\rm R} > \nu_\rmm$ continue to apply
after the specified times. For the narrow component
$\nu_{\rmc,\rmn} > \nu_{\rm R} > \nu_{\rmm,\rmn}$ also
continues to apply since (under the assumptions underlying
eq. [\ref{nu_c2}]) $\nu_{\rmc,\rmn}$ remains constant for
$t> t_{\rmjet,\rmn}$. However, $\nu_{\rmc,\rmw}$ decreases with time after 
$t_{\rmdec,\rmw}$ and could potentially fall below $\nu_{\rm R}$
before its value becomes frozen at $t_{\rmj,\rmw}$. 
The foregoing arguments imply that the frequency regimes
corresponding to equations (\ref{F_narrow_end}),
(\ref{F_wide_4}), (\ref{F_wide_5}), and (\ref{F_wide_end}) would
typically not be relevant.

\clearpage

%\begin{center}
%\epsscale{0.8}
%\plotone{f1.eps}
%\figcaption[]
%{R-band afterglow lightcurve from a two-component jet. The
%contribution of the narrow component, wide component, and their
%sum is represented by the dashed, dash-dotted, and solid curves,
%respectively. The total outflow energy is assumed to be constant,
%$E_\rmw+E_\rmn = 10^{51}\ {\rm ergs}$. From top to bottom, the 3 panels
%correspond to $E_\rmw/E_\rmn= 3$, 1, and 1/3, respectively. The other
%parameters are the same for all panels: 
%$\eta_\rmn = 200$,
%$\eta_\rmw=15$,
%$\theta_{\rmj,\rmw}=0.15$, $\theta_{\rmj,\rmn}=0.05$, $n_0=1$,
%$\epsilon_e=0.1$, $\epsilon_B=0.01$, $p=2.2$, and $D_{\rm L,28}= 1$.
%\label{fig1}}
%\end{center}
%
%\begin{center}
%\epsscale{0.8}
%\plotone{f2.eps}
%\figcaption[]
%{Similar to Fig.~\ref{fig1}, except that $E_\rmw=2\ E_\rmn$ and
%$\theta_{\rmj,\rm w} =2\ \theta_{\rmj,\rmn}$. The top and bottom
%panels correspond to $\theta_{\rmj,\rmn}=0.05$ and 0.1, respectively.
%The other parameters are the same as in Fig.~\ref{fig1}.
%\label{fig2}}
%\end{center}
%
%\begin{center}
%\epsscale{0.8}
%\plotone{f3.eps}
%\figcaption[]
%{R-band lightcurve from a two-component model with parameters
%appropriate to GRB 030329 ($D_{\rm L}=0.8\ {\rm Gpc}$):
%$\eta_\rmn=200$, $\eta_\rmw=8.8$, $E_\rmw=6.4\times 10^{50}\ {\rm ergs}$,
%$E_\rmn=1.8\times 10^{50}\ {\rm ergs}$, $\theta_{\rmj,\rmw}=0.26$,
%$\theta_{\rmj,\rmn}=0.086$, $n_0=0.19\ {\rm cm^{-3}}$,
%$\epsilon_e=0.01$, $\epsilon_B=0.008$, and $p=2.25$.
%\label{fig3}}
%\end{center}
%

\clearpage
\begin{thebibliography}{}
\bibitem[Beloborodov(2000)]{B00}
Beloborodov, A. M. 2000, ApJ, 539, L25

\bibitem[Berger et al.(2003a)]{BKF03}
Berger, E., Kulkarni, S., \& Frail, D. A. 2003a, ApJ, 590, 379

\bibitem[Berger et al.(2003b)]{B03}
Berger, E., et al. 2003b, Nature, 426, 154

\bibitem[Blandford \& McKee(1976)]{BM76}
Blandford, R. D., \& McKee, C. 1976, Phys. Fluids, 19, 1130

\bibitem[Bloom et al.(2003)]{BFK03}
Bloom, J. S., Frail, D. A., \& Kulkarni, S. R. 2003, ApJ, 594,
674

\bibitem[Cannizzo et al.(2004)]{CGV04}
Cannizzo, J. K., Gehrels, N., \& Vishniac, E. T. 2004, ApJ, 601,
380

\bibitem[Fox et al.(2003)]{F03}
Fox, D. W., et al. 2003, Nature, 422, 284

\bibitem[Frail et al.(2000)]{F00}
Frail, D. A., et al. 2000, ApJ, 538, L129

\bibitem[Frail et al.(2001)]{F01}
Frail, D. A., et al. 2001, ApJ, 562, L55

\bibitem[Fynbo et al.(2004)]{F04}
Fynbo, J. P. U., et al. 2004, ApJ, 609, 962

\bibitem[Granot et al.(2001)]{G01}
Granot, J., Miller, M., Piran, T., Suen, W. M., \& Hughes,
P. A. 2001, in Gamma-Ray Bursts in the Afterglow Era,
ed. E. Costa, F. Frontera, \& J. Hjorth (Berlin: Springer), 312

\bibitem[Granot et al.(2003)]{GNP03}
Granot, J., Nakar, E., \& Piran, T. 2003, Nature, 426, 138

\bibitem[Granot et al.(2002)]{G02}
Granot, J., Panaitescu, A., Kumar, P., \& Woosley, S. E. 2002, ApJ, 570, L61

\bibitem[Granot et al.(2005)]{GRL05}
Granot, J., Ramirez-Ruiz, E., \& Loeb, A. 2005, ApJ, 618, 413

\bibitem[Heise(2003)]{H03}
Heise, J. 2003, in AIP Conf. Proc. 662, Gamma-Ray Burst and
Afterglow Astronomy 2001, ed. G. R. Ricker \& R. K. Vanderspek
(Melville: AIP), 229

\bibitem[Hjorth et al.(2003)]{HS03}
Hjorth, J., et al. 2003, Nature, 423, 847

\bibitem[Huang et al.(2004)]{H04}
Huang, Y. F., Wu, X. F., Dai, Z. G., Ma, H. T., \& Lu, T. 2004,
ApJ, 605, 300

\bibitem[Khokhlov et al.(1999)]{K99}
Khokhlov, A. M., H\"oflich, P. A., Oran, E. S., Wheeler, J. C.,
Wang, L., \& Chtchelkanova, Y. 1999, ApJ, 524, L107

\bibitem[Kippen et al.(2003)]{K03}
Kippen, M., Woods, P. M., Heise, J., In't Zand, J. J. M.,
Briggs, M. S., \& Preece, R. D. 2003, in AIP Conf. Proc. 662,
Gamma-Ray Burst and Afterglow Astronomy 2001, ed. G. R. Ricker
\& R. K. Vanderspek (Melville: AIP), 244

\bibitem[Kobayashi \& Sari(2001)]{KS01}
Kobayashi, S., \& Sari, R. 2001, ApJ, 551, 934

\bibitem[Kobayashi \& Zhang(2003)]{KZ03}
Kobayashi, S., \& Zhang, B. 2003, ApJ, 582, L75

\bibitem[K\"onigl(2004)]{K04}
K\"onigl, A. 2004, in AIP Conf. Proc. 727, Gamma-Ray Bursts: 30
Years of Discovery, ed. E. E. Fenimore \& M. Galassi (Melville: AIP), 257

\bibitem[Kumar \& Granot(2003)]{KG03}
Kumar, P., \& Granot, J. 2003, ApJ, 591, 1075

\bibitem[Kumar \& Piran(2000a)]{KP00a}
Kumar, P., \& Piran, T. 2000a, ApJ, 532, 286

\bibitem[Kumar \& Piran(2000b)]{KP00b}
Kumar, P., \& Piran, T. 2000b, ApJ, 535, 152

\bibitem[Lazzati et al.(2002)]{L02}
Lazzati, D., Rossi, E., Covino, S., Ghisellini, G., \& Malesani,
D. 2002, A\&A, 396, L5

\bibitem[Levinson \& Eichler(1993)]{LE93}
Levinson, A., \& Eichler, D. 1993, ApJ, 418, 386

\bibitem[Liang \& Dai(2004)]{LD04}
Liang, E. W., \& Dai, Z. G. 2004, ApJ, 608, L9

\bibitem[Lipkin et al.(2004)]{L04}
Lipkin, Y. M., et al. 2004, ApJ, 606, 381

\bibitem[M\'esz\'aros(2002)]{M02}
M\'esz\'aros, P. 2002, ARAA, 40, 137

\bibitem[Nakar \& Oren(2004)]{NO04}
Nakar, E., \& Oren, Y. 2004, ApJ, 602, L97

\bibitem[Nakar \& Piran(2002)]{NP02}
Nakar, E., \& Piran, T. 2002, ApJ, 572, L139

\bibitem[Nakar \& Piran(2003)]{NP03}
Nakar, E., \& Piran, T. 2003, ApJ, 598, 400

\bibitem[Nakar et al.(2002)]{NPG02}
Nakar, E., Piran, T., \& Granot, J. 2002, ApJ, 579, 699

\bibitem[Nakar et al.(2003)]{NPG03}
Nakar, E., Piran, T., \& Granot, J. 2003, NewA, 8, 495

\bibitem[Panaitescu \& Kumar(2002)]{PK02}
Panaitescu, A., \& Kumar, P. 2002, ApJ, 571, 779

\bibitem[Panaitescu et al.(1998)]{PMR98}
Panaitescu, A., M\'esz\'aros, P., \& Rees, M. J. 1998, ApJ, 503, 314

\bibitem[Pedersen et al.(1998)]{P98}
Pedersen, H., et al. 1998, ApJ, 496, 311

\bibitem[Piran(1999)]{P99}
Piran, T. 1999, Phys. Rep. 314, 575

\bibitem[Piran et al.(2004)]{PNG04}
Piran, T., Nakar, E., \& Granot, J. 2004, in AIP
Conf. Proc. 727, Gamma-Ray Bursts: 30 Years of Discovery,
ed. E. E. Fenimore \& M. Galassi (Melville: AIP), 181

\bibitem[Price et al.(2003)]{P03}
Price, P.A., et al. 2003, Nature, 423, 824

\bibitem[Ramirez-Ruiz et al.(2002)]{RCR02}
Ramirez-Ruiz, E., Celotti, A., \& Rees, M. J. 2002, MNRAS, 337, 1349

\bibitem[Rees \& M\'esz\'aros(1994)]{RM94}
Rees, M.~J., \& M\'esz\'aros, P. 1994, ApJ, 430, L93

\bibitem[Rhoads(1999)]{R99}
Rhoads, J. 1999, ApJ, 525, 737

\bibitem[Rossi et al.(2002)]{RLR02}
Rossi, E., Lazzati, D., \& Rees, M. J. 2002, MNRAS, 332, 945

\bibitem[Sari(1997)]{Sari97}
Sari, R. 1997, ApJ, 489, L37

\bibitem[Sari \& M\'esz\'aros(2000)]{SM00}
Sari, R., \& M\'esz\'aros, P. 2000, ApJ, 535, L33

\bibitem[Sari \& Piran(1995)]{SP95}
Sari, R., \& Piran, T. 1995, ApJ, 455, L143

\bibitem[Sari \& Piran(1999a)]{SP99a}
Sari, R., \& Piran, T. 1999a, ApJ, 520, 641

\bibitem[Sari \& Piran(1999b)]{SP99b}
Sari, R., \& Piran, T. 1999b, A\&A Supl. Ser., 138, 537

\bibitem[Sari et al.(1999)]{SPH99}
Sari, R., Piran, T., \& Halpern J. P., 1999, ApJ, 519, L17

\bibitem[Sari et al.(1998)]{SPN98}
Sari, R., Piran, T., \& Narayan, R. 1998, ApJ, 497, L17

\bibitem[Sheth et al.(2003)]{S03}
Sheth, K., et al. 2003, ApJ, 595, L33

\bibitem[Soderberg et al.(2004)]{S04}
Soderberg, A. M., et al. 2004, ApJ, 606, 994

\bibitem[Starling et al.(2005)]{S05}
Starling, R. L. C., Wijers, R. A. M. J., Hughes, M. A., Tanvir, N. R.,
Vreeswijk, P. M., Rol, E., \& Salamanca, I. 2005, MNRAS, in press
(astro-ph/0501120)

\bibitem[Taylor et al.(2004)]{T04}
Taylor, G.B., Frail, D.A., Berger, E., \& Kulkarni, S.R. 2004, ApJ, 609, L1

\bibitem[Uemura et al.(2003)]{U03}
Uemura, M., Kato, T., Ishioka, R., \& Yamaoka, H. 2003, PASJ, 55L, 31

\bibitem[van Putten \& Levinson(2003)]{VPL03}
van Putten, M. H. P. M., \& Levinson, A. 2003, ApJ, 584, 937

\bibitem[Vlahakis et al.(2003)]{VPK03}
Vlahakis, N., Peng, F., \& K\"onigl, A. 2003, ApJ, 594, L23

\bibitem[Willingale et al.(2004)]{W04}
Willingale, R., Osborne, J. P., O'Brien, P. T., Ward, M. J.,
Levan, A., \& Page, K. L. 2004, MNRAS, 349, 31

\bibitem[Yamazaki et al.(2002)]{Y02}
Yamazaki, R., Ioka, K., \& Nakamura, T. 2002, ApJ, 571, L31

\bibitem[Yamazaki et al.(2004)]{Y04}
Yamazaki, R., Ioka, K., \& Nakamura, T. 2004, ApJ, 606, L33

\bibitem[Yost et al.(2003)]{Y03}
Yost, S. A., Harrison, F. A., Sari, R., \& Frail, D. A. 2003,
ApJ, 597, 459

\bibitem[B. Zhang et al.(2004)]{Z04}
Zhang, B., Dai, X., Lloyd-Ronning, N. M., \& M\'esz\'aros,
P. 2004, ApJ, 601, L119

\bibitem[Zhang \& M\'esz\'aros(2002)]{ZM02}
Zhang, B., \& M\'esz\'aros, P. 2002, ApJ, 571, 876

\bibitem[W. Zhang et al.(2004)]{ZWH04}
Zhang, W., Woosley, S. E. \& Heger, A. 2004, ApJ, 608, 365

\end{thebibliography}
\end{document}